\documentclass[aps,pra,showpacs,twoside,10pt,floatfix,nofootinbib,longbibliography,twocolumn]{revtex4-2}
\usepackage[colorlinks=true, citecolor=red, urlcolor=blue ]{hyperref}
\usepackage{epsfig,newlfont,amssymb,amsfonts,amsmath,bm,subfigure,palatino,mathtools,amsthm,braket,soul,enumitem,color,times,comment,geometry,xcolor}
\usepackage[normalem]{ulem}
\definecolor{checkcolor}{HTML}{7541C0}

\begin{document}

\title{Quantum error correction via purification using a single auxiliary}
\author{Chandrima B. Pushpan$^1$, Tanoy Kanti Konar$^2$, Aditi Sen(De)$^2$, Amit Kumar Pal$^1$}
\affiliation{$^1$Department of Physics, Indian Institute of Technology Palakkad, Palakkad 678 623, India \\
$^2$Harish-Chandra Research Institute, A CI of Homi Bhabha National Institute, Chhatnag Road, Jhunsi, Prayagraj 211 019, India}

\begin{abstract}

We propose a single  auxiliary-assisted purification-based framework for quantum error correction, capable of correcting errors that drive a system from its ground-state subspace into excited-state sectors. The protocol consists of a joint time evolution of the system-auxiliary duo under a specially engineered interaction Hamiltonian, followed by a single measurement of the auxiliary in its energy eigenbasis and a subsequent post-selection of one of the measurement outcomes. We show that the resulting purified state always achieves unit fidelity, while the probability of obtaining any energy of the auxiliary other than its ground state energy yields the success rate of the protocol. We demonstrate the power of this proposed method for several low-distance quantum codes, including the three-, four-, and five-qubit codes, and for the one-dimensional isotropic Heisenberg model, subjected to bit-flip, phase-flip, and amplitude-damping noises acting on all qubits. Notably, the protocol expands the class of correctable errors for a given code, particularly in the presence of amplitude-damping noise. We further analyze the impact of replacing the auxiliary qudit with a single auxiliary qubit, and the changes in the performance of the protocol under the realistic
scenario where noise remains active during the correction cycle.

\end{abstract}

\maketitle

\section{Introduction}
\label{sec:intro}

The world-wide quest of developing a fault-tolerant scalable quantum computer~\cite{Shor1996a,preskill1998,Gottesman2013} in order for solving complex physical problems faster than existing classical computers~\cite{Shor1997,harrow2017} has led to an ecosystem of quantum error handling techniques. On one hand, quantum error correction (QEC)~\cite{Shor1996a,preskill1998,Gottesman2013,nielsen2010,Terhal2015,Roffe2019} provides a systematic road map to protect information against decoherence by detecting and correcting errors without disturbing the underlying resource states. On the other hand, quantum error mitigation~\cite{Temme2016,Endo2018,Suzuki2022}, developed primarily for the noisy intermediate-scale quantum (NISQ) devices~\cite{Preskill_quantum_2018_nisq,nash2020}, is used to reduce the effect of noise on the output of quantum computation without running a full QEC protocol. Further, regulating spreading of errors during logical operations is attempted via magic state distillation~\cite{Bravyi2005}, fault-tolerant transversal gate construction~\cite{Eastin2009}, code switching~\cite{Paetznick2013}, and lattice surgery~\cite{Horsman2011}, while reducing the error rates via hardware-level noise suppression is achieved by the dynamical decoupling~\cite{Viola1998}, and by using appropriate continuous-variable quantum codes such as the bosonic Schr\"{o}dinger cat encoding~\cite{Cochrane1998} and the Gottesman-Kitaev-Preskill (GKP) code~\cite{Gottesman2000}.

Arguably, the most extensively studied approach, and therefore the most mature among these techniques is the QEC~\cite{Shor1996a,preskill1998,Gottesman2013,nielsen2010,Terhal2015,Roffe2019}. In this approach, information is encoded in the low-energy \emph{logical} subspace of a large Hilbert space associated to a multi-qubit system referred to as the quantum error-correcting code (QECC)~\cite{nielsen2010,Terhal2015,Roffe2019}. 
Specifically, the physical qubits are positioned on lattices with  specific geometries, and interact 
with each other via engineered multi-qubit interactions.  In a quantum protocol, the QECC is prepared in a \emph{fiducial state} belonging to the logical subspace, while errors originating from different sources take the QECC out of the logical subspace to a state with higher energy. A QEC procedure is said to be successful  when it identifies and corrects these errors, returning the system to the logical subspace. 

A wide variety of QECCs have emerged to address diverse error models and operational requirements. Among the low-distance (\(d \leq 3\)) codes in the Calderbank–Shor–Steane (CSS) family~\cite{Calderbank1996} are the seven-qubit Stean code~\cite{Steane1996,Steane1996_a} and the nine-qubit Shor code~\cite{Shor1995}. More recently, tunable-distance topological QECCs~\cite{fujii2015,*prakash_pra_2015}, including the Kitaev surface code~\cite{kitaev2001,*kitaev2003,*kitaev2006} and the color code~\cite{bombin2006,*bombin2007}, have attracted lots of interest. Additionally, several non-CSS low-distance codes have been developed, such as the three-qubit bit-flip code~\cite{nielsen2010,Terhal2015,Roffe2019,Shor1995}, the four-qubit Grassl–Beth–Pellizzari code~\cite{nielsen2010,Terhal2015,Roffe2019,Grassl1997}, and the five-qubit perfect code~\cite{nielsen2010,Terhal2015,Roffe2019,Laflamme1996}.
In the last two  decades, realizations of QECCs in trapped ions~\cite{haffner2008} and superconducting qubits~\cite{Krantz2019,Kjaergaard2020,Bravyi2022} have enabled researchers to perform proof-of-principle experiments with small QECCs~\cite{Chiaverini2004,*Schindler2011,*Nigg2014,*Linke2017,*Pal_quantum_2022_relaxation,Reed2012,*Kelly2015,*Andersen2020,*Zhao2022,*Krinner2022May} as well as explore fault-tolerant scalable quantum computation~\cite{Acharya2024,gao2024}.

Besides detecting error via syndrome measurements followed by a correction protocol, QEC have also been approached from different viewpoints. The purification of entangled states has been connected to QEC~\cite{Bennett1996_MSEC}. Considering the errors as sources of entropy in the system,  QEC has also been interpreted as a continuous cooling mechanism implemented via engineered nonlinear dissipation~\cite{Sarovar2005,Sarovar2005_a,Leghtas2015,Hillmann2023} (cf.~\cite{Diehl2008}).
It has also been  explored as appropriate engineering of system-bath couplings to drive the system  through open system dynamics to a desired logical subspace as its steady state~\cite{Kraus2008,Verstraete2009}.  Autonomous QEC uses continuous, engineered interactions to autonomously and continuously correct errors without external measurement and feedback~\cite{Kerckhoff2010,Cohen2014,Reiter2017,Touzard2018,Li2024}. Further, while QEC has been explored as a quantum refrigeration~\cite{Ben-Or2013} without the presence of any auxiliary,  recently, auxiliary assisted constructions, referred to as quantum information-theoretic devices, capable of correcting arbitrary noise via purification in an $N$-qubit Clifford circuit, have also been proposed~\cite{Das2024}, using $2N$ auxiliary qubits. 


In this work, we introduce a \emph{single auxiliary-assisted} QEC framework, in which transitions from the different degenerate energy subspaces of the Hilbert space associated to a quantum system of interest to its ground state are induced by an engineered interaction of the system with an auxiliary qudit. We specifically examine a scenario where the fiducial state of the system, belonging typically to the ground state subspace, is subjected to errors resulting in a mixed state. Upon turning on the interaction between the system initialized in this mixed state and the auxiliary in its ground state at $t=0$, 
the subsequent time evolution followed by a suitable measurement on the auxiliary purifies the logical state with unit fidelity and a finite probability of success.
We apply this methodology to several small QECCs including the three-qubit bit-flip code, the four-qubit Grassl code, and the five-qubit perfect code subjected to the bit-flip, the phase-flip, and the amplitude-damping errors, and demonstrate successful purification of logical states with finite probabilities. In this context, we point out that correcting amplitude-damping error is a challenging task~\cite{cafao_2014,grassl_2018,dutta_2025,dutta_2025_arxiv,wang_2025}. Our proposal, however, is successful in handling amplitude-damping error even in QECCs that are unable to correct amplitude-damping error using the conventional QEC protocols.

We also show the applicability of our protocol
in purifying ground states of one-dimensional isotropic Heisenberg model~\cite{giamarchi2004,Franchini2017}, which is used as  logical qubits in quantum state transfer in two-dimensional rectangular lattices~\cite{Pushpan2024,Pushpan2024Jul}. While the dimension of the auxiliary system in the proposed framework is determined by the number of degenerate energy subspaces of the system,  we explore the pertinent question of whether reducing this resource impacts performance. 
Towards this, we construct specific interaction Hamiltonians that connect the different energy subspaces of the system with a single qubit, and demonstrate that the proposed protocol still achieves unit fidelity with finite success probability. 
We also analyze the realistic scenarios where the noise on the logical subspace keep acting on the system, and discuss how the outcome of the protocol changes under these conditions.    

The rest of this paper is organized as follows. In Sec.~\ref{sec:Framework}, we introduce the mathematical framework of QEC via purification using a single auxiliary qudit. In Sec.~\ref{sec:common_errors}, we apply the devised methodology to low-distance QECCs including the three-  (Sec.~\ref{subsec:three_qubit}), four- (Sec.~\ref{subsec:four_qubit}), and five-qubit (Sec.~\ref{subsec:five_qubit}) codes subjected to bit-flip, phase-flip, and amplitude-damping errors. We discuss the use of the protocol in the ground-state purification of the 1D isotropic Heisenberg model in Sec.~\ref{sec:heisenberg_model}, and the effect of replacing the auxiliary qudit with a qubit in Sec.~\ref{sec:purification_with_qubit}. Sec.~\ref{sec:resilient_noise} elaborates on the performance of the protocol in the realistic scenario where the noise can be resilient. Sec.~\ref{sec:conclusion} includes the concluding remarks.

\section{Formalism: Correction via purification}
\label{sec:Framework}

\begin{figure*}
    \centering
    \includegraphics[width=0.8\linewidth]{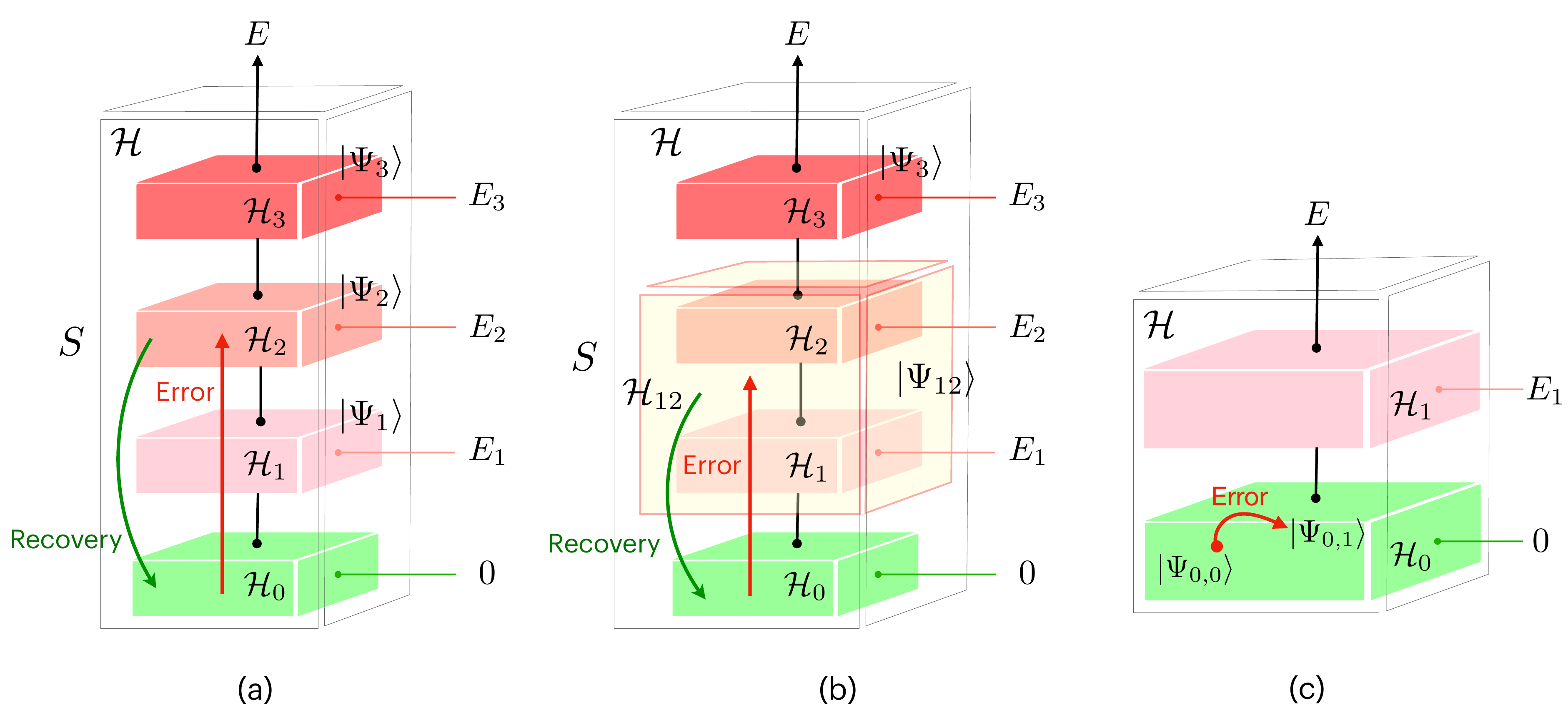}
    \caption{(a) Splitting of $\mathcal{H}$ into the orthogonal degenerate energy subspaces $\mathcal{H}_i$ having energy $E_i$, such that $\mathcal{H}=\oplus_{i=0}^D\mathcal{H}_i$. An error corresponds to taking the system out of $\mathcal{H}_0$, while a recovery takes it back to $\mathcal{H}_0$. (b) There may exist errors that take the system out of $\mathcal{H}_0$ so that non-zero overlap can be found with more than one $\mathcal{H}_i$, eg. $\mathcal{H}_1$ and $\mathcal{H}_2$. An archetype state $\ket{\Psi_{12}}$ constructed in this case has non-zero overlap with both $\mathcal{H}_1$ and $\mathcal{H}_2$. (c) An uncorrectable error in our framework corresponds to the error that keeps the system within the ground state subspace $\mathcal{H}_0$.}
    \label{fig:schematic}
\end{figure*}

Consider a Hamiltonian $H_S$ describing a $D_S$-dimensional quantum system, satisfying the eigenvalue equation
\begin{eqnarray}
H_S\ket{E_{i,\alpha}}=E_i\ket{E_{i,\alpha}},
\end{eqnarray}
where $i=0,1,2,\cdots,D$ ($1\leq D\ll D_S$), and $\alpha=0,1,2,\cdots,d_i-1$, with $E_i$ being $d_i$-fold degenerate, such that the subspaces 
\begin{eqnarray}
    \mathcal{H}_i=\text{span}\set{\ket{E_{i,0}},\ket{E_{i,1}},\cdots,\ket{E_{i,d_i-1}}}
\end{eqnarray}
constitute the full Hilbert space $\mathcal{H}$ as $\mathcal{H}=\oplus_{i=0}^{D}\mathcal{H}_i$. Here, we assume $\langle E_{i,\alpha}|E_{j,\beta}\rangle=0$ for $i\neq j$ irrespective of $\alpha$ and $\beta$, and $\langle E_{i,\alpha}|E_{i,\beta}\rangle=0$ for $\alpha\neq \beta$, while we set $E_0=0$ (see Fig.~\ref{fig:schematic}(a)).

We consider a particular orthonormal set of states spanning $\mathcal{H}_0$, given by 
\begin{eqnarray}
    \ket{\Psi_{0,\beta}}=\sum_{\alpha=0}^{d_0-1}a_{\alpha,\beta}\ket{E_{0,\alpha}}, 
    \label{eq:resource}
\end{eqnarray}
with $\beta=0,1,2,\cdots,d_0-1$, to serve as resource for a specific quantum information processing task, and assume that an error $\mathcal{E}$ on one of the resource state, say, $\ket{\Psi_{0,\gamma}}$, resulting in 
\begin{eqnarray}
    \ket{\Psi_{0,\gamma}}\rightarrow\ket{\Phi_i}=\mathcal{E}\ket{\Psi_{0,\gamma}}\in\mathcal{H}_i, i\neq 0
\end{eqnarray}
takes the system out of $\mathcal{H}_0$ to $\mathcal{H}_i$, causing an increase  in the energy by $E_i$. We consider this to be a \emph{selective} error as it projects the system onto one specific energy subspace $\mathcal{H}_i$, as opposed to the situation where the error is \emph{non-selective}, leading to a state having overlap with multiple energy subspaces.   

\paragraph{Strategy for selective errors.} A correction $\mathcal{E}^{-1}$ to this selective error, given by 
\begin{eqnarray}
    \ket{\Phi_i}\rightarrow\ket{\Psi_{0,\gamma}}=\mathcal{E}^{-1}\ket{\Phi_i},
\end{eqnarray}
requires the energy $E_i$ to dissipate on an auxiliary system $A$ of dimension $D+1$ (a qudit), described by the Hamiltonian
\begin{eqnarray}
    H_{A}=\sum_{i=0}^{D}E_i\ket{e_i}\bra{e_i}, 
\end{eqnarray}
initialized in its zero-energy ($E_0=0$) ground state $\ket{e_0}$, resulting in a jump in the energy by $E_i$ to an excited state $\ket{e_i}$. While this can be typically achieved by a unitary operation resulting in $\ket{\Psi_i}\ket{e_0}\leftrightarrow \ket{\Psi_{0,\gamma}}\ket{e_i}$, in this paper, we achieve this using a time-evolution generated by the Hamiltonian that results in energy-conserving transitions between the states $\ket{\Psi_i}\ket{e_0}$ and $\ket{\Psi_{0,\gamma}}\ket{e_i}$ of the joint system $SA$, given by 
\begin{eqnarray}
    H_{SA}^{(i)}=g_i(\ket{\Psi_{0,\gamma}e_i}\bra{\Psi_i e_0}+\text{h.c.}).
    \label{eq:interaction_SA}
\end{eqnarray}
Here, $g_i$ is the strength of the interaction between the system and the auxiliary, and $\ket{\Psi_i}$ is an archetype state $\in\mathcal{H}_i$, which we construct, and has the form
\begin{eqnarray}
    \label{eq:archetype_states}
    \ket{\Psi_i}&=&\left\{\begin{array}{ll}
       \sum_{\alpha=0}^{d_i-1}b_\alpha \ket{E_{i,\alpha}}  &\text{ for }d_i>1,  \\
       \ket{E_{i,0}}  &\text{ for }d_i=1, 
    \end{array}\right.
\end{eqnarray}
with judiciously choosing $b_{\alpha}$ such that $b_\alpha$ are real and $\sum_{\alpha=0}^{d_i-1}b_\alpha^2=1$ (we confine ourselves to real $b_\alpha$ in order to keep $H_{SA}^{(i)}$ real). 
Note that we do not assume any detail regarding the error $\mathcal{E}$ except the transition in energy it causes, which refrains us from modeling $H_{SA}^{(i)}$ using the state $\ket{\Phi_i}$ itself. Note further that the dynamics generated by this Hamiltonian is confined within the subspaces $\mathcal{H}_0$ and $\mathcal{H}_i$, while all other subspaces $\mathcal{H}_{j\neq 0}$ and $\mathcal{H}_{j\neq i}$  remain immune to it. 

Decomposing the system and the auxiliary Hamiltonians as $H_S=H_S^{(i)}+H_S^{\text{rest}}$ and $H_A=H_A^{(i)}+H_A^{\text{rest}}$ with
\begin{eqnarray}
    H_S^{(i)} &=& E_i\sum_{\alpha=0}^{d_i-1}\ket{E_{i,\alpha}}\bra{E_{i,\alpha}},\nonumber\\
    H_S^{\text{rest}} &=& \sum_{j\neq 0,i} E_j\sum_{\alpha=0}^{d_j-1}\ket{E_{j,\alpha}}\bra{E_{j,\alpha}},
\end{eqnarray}
and 
\begin{eqnarray}
    H_A^{(i)}&=& E_i\ket{e_i}\bra{e_i},\nonumber\\
    H_A^{\text{rest}}&=&\sum_{j\neq 0,i}E_j\ket{e_j}\bra{e_j},
\end{eqnarray}
and noticing that $[H_S^{\text{rest}},H_S^{(i)}]=[H_A^{\text{rest}},H_A^{(i)}]=0$, and $[H_{X}^{\text{rest}},H_{SA}^{(i)}]=0$ ($X=S,A$),  one can identify $H^{(i)}=H_S^{(i)}+H_A^{(i)}+H_{SA}^{(i)}$ to generate the non-trivial part of the time-evolution as $U^{(i)}=\text{e}^{-\text{i}H^{(i)}t}$. Writing 
\begin{eqnarray}
    \ket{E_{0,\alpha}}=\sum_{\beta=0}^{d_0-1}a^\prime_{\alpha,\beta}\ket{\Psi_{0,\beta}},
\end{eqnarray}
with $\sum_{\beta=0}^{d_0-1}(a^\prime_{\alpha\beta})^2=1$, one can obtain 
\begin{eqnarray}
    \text{e}^{-\text{i}H^{(i)}t}\ket{\Phi_ie_0} &=& \sqrt{P}\ket{\Psi_{0,\gamma}e_i}+\cdots,
    \label{eq:time_evolution}
\end{eqnarray}
where $P$ depends on (a) the specific choice of the archetype state $\ket{\Psi_i}$, (b) the value of $g_i$, and (c) the coefficients $a_{\alpha,\gamma}$. Performing a measurement subsequently on the auxiliary system in its energy eigenbasis $\{\ket{e_i};\;i=0,1,\cdots,D\}$ collapses $S$ to $\ket{\Psi_{0,\gamma}}$ with a probability $P$. Therefore, a time-evolution of $\ket{\Phi_i e_0}$ with $H^{(i)}$ followed by a suitable measurement of $A$ returns the resource state $\ket{\Psi_{0,\gamma}}$ with a success rate $P$ (cf. \cite{Nakazato_2003, Nakazato_2004}).   

\paragraph{Elimination policy for non-selective error.} When an error $\mathcal{E}$ is non-selective, i.e., takes $\ket{\Psi_{0,\gamma}}$ to a state $\ket{\Phi}$ having overlap with not one, but a subset $\mathcal{H}=\mathcal{H}_{i_1}\cup\mathcal{H}_{i_2}\cup\cdots\cup\mathcal{H}_{i_r}$ of energy subspaces, where $\{i_1,i_2,\cdots,i_r\}\subset\{1,2,\cdots,D\}$. Note that the above framework is flexible to combine all such energy subspaces, and constitute a normalized archetype state as 
\begin{eqnarray}
    \ket{\Psi_{i_1,\cdots,i_r}}= \sum_{i=i_1}^{i_r}\sum_{\alpha_i=0}^{d_i-1}b_{i,\alpha_i}\ket{E_{i,\alpha_i}},
    \label{eq:combined_subspaces}
\end{eqnarray}
and subsequently include 
\begin{eqnarray}
    H_{SA}^{(i_1,\cdots,i_r)}&=&g_{i,j}(\ket{\Psi_{0,\gamma}e_k}\bra{\Psi_{i_1,\cdots,i_r} e_0}+\text{h.c.}),
    \label{eq:combined_subspace_interaction}
\end{eqnarray}
in $H_{SA}$, where the energy corresponding to $\ket{e_k}$ in $A$ can be set to $\tilde{E}_k=\langle \Psi_{i_1,\cdots,i_r}|H_S|\Psi_{i_1,\cdots,i_r}\rangle$ (see Fig.~\ref{fig:schematic}(b) for an example with $\mathcal{H}_1$ and $\mathcal{H}_2$).

\paragraph{General framework.} In general, error rates $q$ ($0\leq q\leq 1$) are associated to each selective errors $\mathcal{E}$,  such that the corresponding noise model $\Lambda(.)$ results in 
\begin{eqnarray}
    \ket{\Psi_{0,\gamma}}&\rightarrow&\rho_S^0\nonumber\\&=&\Lambda(\ket{\Psi_{0,\gamma}}\bra{\Psi_{0,\gamma}}),\nonumber\\
    &=&(1-q)\ket{\Psi_{0,\gamma}}\bra{\Psi_{0,\gamma}}+q\mathcal{E}\ket{\Psi_{0,\gamma}}\bra{\Psi_{0,\gamma}}\mathcal{E}^{\dagger},\nonumber\\
\end{eqnarray}
where $\rho_S^0$ is typically a mixed state. The general framework discussed above works also in this situation, with 
\begin{eqnarray}
    \rho_{SA}(t)&=&U^{(i)}\rho_S^0\otimes\ket{e_0}\bra{e_0}U^{(i)\dagger}\nonumber\\&=&P\ket{\Psi_{0,\gamma}}\bra{\Psi_{0,\gamma}}\otimes\ket{e_i}\bra{e_i}+\cdots,
    \label{eq:time_evolution_density_matrix_form}
\end{eqnarray}
where, in addition to the factors mentioned above (see Eq.~(\ref{eq:time_evolution})), $P$ is a function of $q$ also, and the framework serves as a purification protocol. In the most general situation where multiple energy subspaces are involved and different rates are associated to different errors, $\rho_S^0$ can be written as 
\begin{eqnarray}
    \rho_{S}^0=\sum_{i=0}^{D}\sum_{\alpha=0}^{d_i-1}\sum_{j=0}^{D}\sum_{\beta=0}^{d_j-1}\rho_{i,j,\alpha,\beta}\ket{E_{i,\alpha}}\bra{E_{j,\beta}},  
\end{eqnarray}
and one may exploit all $(D+1)$ levels of the qudit via generalizing the system-auxiliary interaction Hamiltonian as $H_{SA}=\sum_{i=1}^{D}H_{SA}^{(i)}$, with $H_{SA}^{(i)}$ given in Eq.~(\ref{eq:interaction_SA}), such that the time evolution $U=\text{e}^{-\text{i}H t}$, generated by the total system-auxiliary Hamiltonian $H=H_S+H_A+H_{SA}$, incorporates contributions of all energy subspaces in  the success rate $P$. Due to this, Eq.~(\ref{eq:time_evolution_density_matrix_form}) modifies to 
\begin{eqnarray}
    \rho_{SA}(t)&=&U\rho_S^0\otimes\ket{e_0}\bra{e_0}U^\dagger\nonumber\\&=&\ket{\Psi_{0,\gamma}}\bra{\Psi_{0,\gamma}}\otimes\sum_{i=1}^{D}\tilde{P}_i\ket{e_i}\bra{e_i}+\cdots,
    \label{eq:most_general_situation_closed}
\end{eqnarray}
implying that each measurement outcome resulting in a collapse of the qudit to the state $\ket{e_i}$ corresponds to a collapse of the system to the state $\ket{\Psi_{0,\gamma}}$ with probability $\tilde{P}_i$.  The \emph{total} success rate of the protocol is $P=\sum_{i=1}^D\tilde{P}_i$. 

\paragraph{Example.} As the simplest example, the above framework allows for purifying an arbitrary single-qubit state 
\begin{eqnarray}
    \rho_{S}^0=r\ket{0}\bra{0}+(1-r)\ket{1}\bra{1}+(s\ket{0}\bra{1}+\text{h.c.}),
\end{eqnarray}
with $r$ ($s$) real (complex), to its ground state $\ket{0}$. Using $H_{SA}$ of the form
\begin{eqnarray}
    H_{SA}=g\left(\ket{0e_1}\bra{1e_0}+\ket{1e_0}\bra{0e_1}\right), 
\end{eqnarray}
one obtains
\begin{eqnarray}
    \rho_{SA}(t)&=&(1-r)(\cos^2gt\ket{1e_0}\bra{1e_0}+\sin^2gt\ket{0e_1}\bra{0e_1})\nonumber \\
    &&+r\ket{0e_0}\bra{0e_0}+\big[\{\text{i}(1-r)/2\}\ket{1e_0}\bra{0e_1}\nonumber \\
    &&+s\text{e}^{\text{i}E_1t}(\cos gt\ket{0e_0}\bra{1e_0}+\text{i}\sin gt \ket{0e_0}\bra{0e_1})\nonumber \\
    &&+\text{h.c.}\big],
\end{eqnarray}
leading to 
\begin{eqnarray}
    P&=&\sin^2 gt (1-r). 
\end{eqnarray}
For a given $r$, $P$ is maximum at $t=(2\ell+1)\pi/2g$, $\ell=0,1,2,\cdots$, $\forall g$, with $P_{\max}=1-r$, the population of the excited state of the qubit in $\rho_S^0$. Further, $P_{\max}\neq 0$ as long as $r\neq 1$, with $r=1$ corresponding to a situation where the entire population is in the ground state.

\noindent\textbf{Remark.} It is important to note that the design of the protocol allows for the success rate $P$ to be the probability with which any one of $\set{e_i}$ except $e_0$ is obtained as the measurement outcome, while the failure probability is the probability of obtaining $e_0$, given by $1-P$. Therefore, in principle, any Hermitian operator on the Hilbert space of $A$ can be chosen for this measurement as long as $\ket{e_0}$ is one of the eigenvectors of it. Note also that the protocol is oblivious to an error $\mathcal{E}$ that takes $\ket{\Psi_{0,\gamma}}$ to a state in $\mathcal{H}_0$ (see Fig.~\ref{fig:schematic}(c)).

\begin{figure*}
    \centering
    \includegraphics[width=0.8\linewidth]{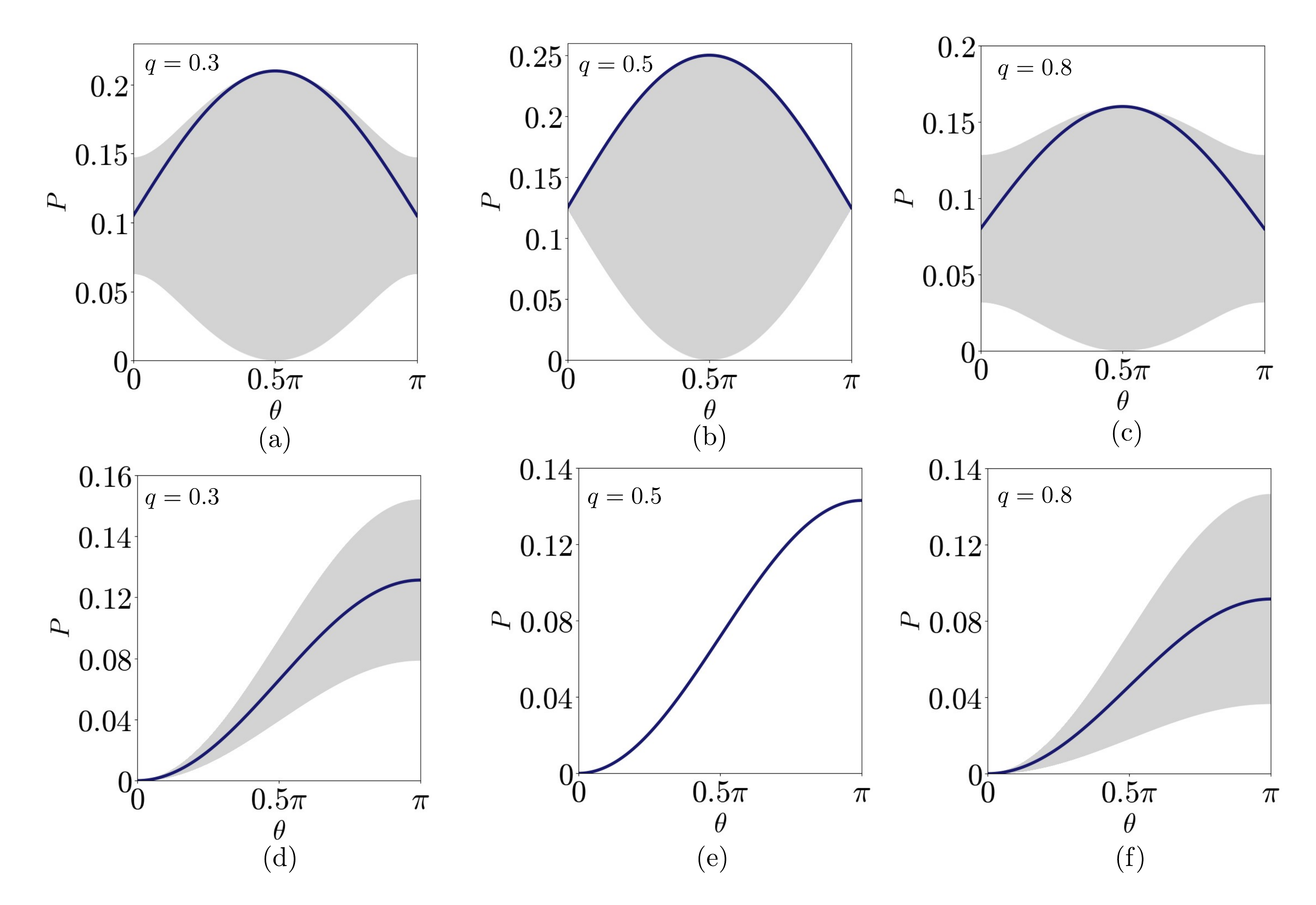}
    \caption{\textbf{Three-qubit code.} Variations of $P$ (ordinate) as a function of $\theta$ (abscissa) for different error rates $q$ when an arbitrary state in the logical subspace is subjected to (a)-(c) bit-flip, and  (d)-(f) amplitude-damping errors on all qubits. The grey lines correspond to randomly sampled $\ket{\Psi_i}\in\mathcal{H}_i$, while the dark continuous line corresponds to the choice of equal superposition of contributing $\ket{E_{1,\alpha}}\in\mathcal{H}_1$ in $P$.  The horizontal axes in all figures are in radian, and $t=\pi/2g$ for all plots.}
    \label{fig:3qubit}
\end{figure*}



\section{Correcting common errors in low-distance quantum codes}
\label{sec:common_errors}

Let us now demonstrate the power of the above framework for a number of systems affected by the well-known noise, namely, the bit-flip (BF), the phase-flip (PF), and the amplitude-damping (AD) noise. We follow the Krauss operator formalism~\cite{nielsen2010,Petruccione,rivas2012,Lidar2019} to determine the effect of the noise on the system. Assuming that the $N$-qubit system is prepared in the resource state $\varrho_S^0$ (typically a state in $\mathcal{H}_0$),  application of a specific error (BF, PF, or AD) with error-rate $q_i$  on the qubit $i$ ($i=1,2,\cdots,N$) leads to the state 
\begin{eqnarray}
    \rho_S^0=\sum_{k=0}^{2^N-1}K_k\varrho_S^0 K_k^\dagger,
    \label{eq:noisy_logical_state}
\end{eqnarray}
where $K_{k}=\otimes_{i=1}^NK_{k_i}$ with $k_i\in\{0,1\}$, $k$ is the decimal representation of the bit-string $k_1k_2k_3\cdots k_N$, and $K_{k_i}$ are the  Krauss operators. For the BF and the PF noises, $K_{k_i}$ are given by~\cite{nielsen2010,Petruccione,rivas2012,Lidar2019} 
\begin{eqnarray}
    \label{eq:bit_flip_krauss}\text{BF}&:&K_{k_i=0}=\sqrt{1-q_i}I, K_{k_i=1}=\sqrt{q_i}\sigma^x,\\
    \label{eq:phase_flip_krauss}\text{PF}&:&K_{k_i=0}=\sqrt{1-q_i}I, K_{k_i=1}=\sqrt{q_i}\sigma^z,
\end{eqnarray}
while in the case of the AD noise,
\begin{equation}
\label{eq:amplitude_damping_krauss}
K_{k_i=0} =
\begin{bmatrix}
1 & 0 \\
0 & \sqrt{1-q_i}
\end{bmatrix},
\quad
K_{k_i=1} =
\begin{bmatrix}
0 & \sqrt{q_i} \\
0 & 0
\end{bmatrix}.
\end{equation}

Let us first consider the situations where the sources of these noises are removed after a finite time, when one is left with a faulty (mixed) state due to these noises, and uses the proposed protocol to get the original resource state back. As systems, we first consider small (low-distance) quantum codes. Among the large collection of such codes, we assess the effectiveness of our framework in the three-qubit repetition code~\cite{nielsen2010,Terhal2015,Roffe2019,Shor1995}, the four-qubit Grassl code~\cite{nielsen2010,Terhal2015,Roffe2019,Grassl1997}, and the five-qubit perfect code~\cite{nielsen2010,Terhal2015,Roffe2019,Laflamme1996}. We assume that all qubits are affected by the same error (i.e., either BF, or PF, or AD) with the same error rate, i.e., $q_i=q$ $\forall i=1,2,\cdots,N$.

\begin{table*}[]
    \centering
    \begin{tabular}{|l|l|l|l|}
    \hline 
    Subspaces & $\mathcal{H}_1$ & $\mathcal{H}_2$ & $\mathcal{H}_3$ \\
    \hline 
    Spanning elements &
    \begin{tabular}{l}
    $\ket{E_{1,0}}=-i\sigma^x_3\sigma^y_4\ket{E_{0,0}}$\\
    $\ket{E_{1,1}}=\sigma^x_4\ket{E_{0,0}}$\\
    $\ket{E_{1,2}}=\sigma^x_2\ket{E_{0,0}}$\\
    $\ket{E_{1,3}}=\sigma^x_3\ket{E_{0,0}}$\\
    $\ket{E_{1,4}}=\sigma^x_2\ket{E_{0,1}}$\\
    $\ket{E_{1,5}}=\sigma^z_3\ket{E_{0,1}}$ \\
    \end{tabular}     &  
    \begin{tabular}{l}
    $\ket{E_{2,0}}=\sigma^x_3\sigma^z_4\ket{E_{0,1}}$\\
    $\ket{E_{2,1}}=\sigma^x_4\sigma^z_3\ket{E_{0,1}}$\\
    $\ket{E_{2,2}}=i\sigma^y_2\ket{E_{0,0}}$\\
    $\ket{E_{2,3}}=\sigma^x_2\sigma^x_3\ket{E_{0,1}}$\\
    $\ket{E_{2,4}}=\sigma^x_2\sigma^x_3\ket{E_{0,0}}$\\
    $\ket{E_{2,5}}=\sigma^x_2\sigma^z_3\ket{E_{0,1}}$\\
    \end{tabular}     &
    \begin{tabular}{l}
    $\ket{E_{3,0}}=i\sigma^y_2\sigma^x_3\ket{E_{0,1}}$\\
    $\ket{E_{3,1}}=i\sigma^y_2\sigma^x_3\ket{E_{0,0}}$
    \end{tabular}\\
    \hline 
    \end{tabular}
    \caption{Elements of the energy eigenbasis of the four-qubit Grassl code that span the different subspaces $\mathcal{H}_i$, $i=1,2,3$.}
    \label{tab:4_qubit}
\end{table*}

\subsection{Three-qubit repetition code} 
\label{subsec:three_qubit}

Let us first consider the three-qubit $(N=3)$ stabilizer Hamiltonian,
\begin{eqnarray}
    H_S=-J(\sigma^z_1\sigma^z_2+\sigma^z_2\sigma^z_3+\sigma^z_3\sigma^z_1)+3JI,
\end{eqnarray}
constituted of the stabilizers $S_1=\sigma^z_1\sigma^z_2$, $S_2=\sigma^z_2\sigma^z_3$, and $S_3=\sigma^z_1\sigma^z_3$ of the three-qubit repetition code capable of correcting single-qubit bit-flip errors in conventional quantum error correction schemes~\cite{nielsen2010,Terhal2015,Roffe2019,Shor1995}. Here, $J$ is the strength of interaction,  $\sigma^\alpha$, $\alpha=x,y,z$ are the Pauli matrices, and $I$ is the identity operator on the Hilbert space of the relevant system. The Hamiltonian $H_S$ exhibits a two-fold degenerate ground state subspace spanned by 
\begin{eqnarray}
    \{\ket{E_{0,0}}=\ket{000},\ket{E_{0,1}}=\ket{111}\},
\end{eqnarray}
corresponding to the zero-energy states, and a six-fold degenerate excited state subspace spanned by
\begin{eqnarray}
    \{\ket{E_{1,0}}&=&\ket{100}, \ket{E_{1,1}}=\ket{010}, \ket{E_{1,2}}=\ket{001},\nonumber \\
    \ket{E_{1,3}}&=&\ket{011}, \ket{E_{1,4}}=\ket{110}, \ket{E_{1,5}}=\ket{101}\},
\end{eqnarray}
with energy eigenvalue $4J$. A generic state  within the ground-state subspace, along with its orthogonal counterpart, is given by
\begin{eqnarray}
\ket{\Psi_{0,0}}&=&\cos\frac{\theta}{2}\ket{000}+\text{e}^{\text{i}\phi}\sin\frac{\theta}{2}\ket{111},\nonumber\\
\ket{\Psi_{0,1}}&=&\sin\frac{\theta}{2}\ket{000}-\text{e}^{\text{i}\phi}\cos\frac{\theta}{2}\ket{111},
\label{eq:ground_states}
\end{eqnarray}
with $0\leq \theta\leq \pi$ and $0\leq \phi\leq 2\pi$. We assume $\varrho_S^0=\ket{\Psi_{0,0}}\bra{\Psi_{0,0}}$. Given that $D=1$, we consider an auxiliary qubit ($D+1=2$), defined by the Hamiltonian $H_A=E_1\ket{1}\bra{1}$, and construct $H_{SA}$  using $\ket{\Psi_1}$ having the form as given in Eq.~(\ref{eq:archetype_states}).

\paragraph{Bit-flip error.}When local BF error occurs on all the qubits of $\varrho_S^0$ (Eq.~(\ref{eq:noisy_logical_state})). The time evolution in the presence of the auxiliary (see Eq.~(\ref{eq:most_general_situation_closed})) followed by a measurement of the auxiliary and post-selection of the measurement outcome $e_1$, leads to the success rate 
\begin{eqnarray}
    P&=&\sin^2 gt\big[A_1q(1-q)^2+A_2q^2(1-q)\big].
    \label{eq:P_bit_flip}
\end{eqnarray}
with $A_1=\sum_{\alpha=0}^2U(b_\alpha,b_{\alpha+3})$ and $A_2=\sum_{\alpha=0}^2V(b_\alpha,b_{\alpha+3})$, 
where 
\begin{eqnarray}
U_{\pm}(x,y)&=&x^2 \cos^2\frac{\theta}{2}\pm xy\sin{\theta}\cos{\phi}+y^2 \sin^2\frac{\theta}{2},\nonumber\\
V(x,y)&=&x^2 \sin^2\frac{\theta}{2}+xy\sin{\theta}\cos{\phi}+y^2 \cos^2\frac{\theta}{2}.
\label{eq:UV_definitions}
\end{eqnarray}

\begin{figure*}
    \centering
    \includegraphics[width=0.8\linewidth]{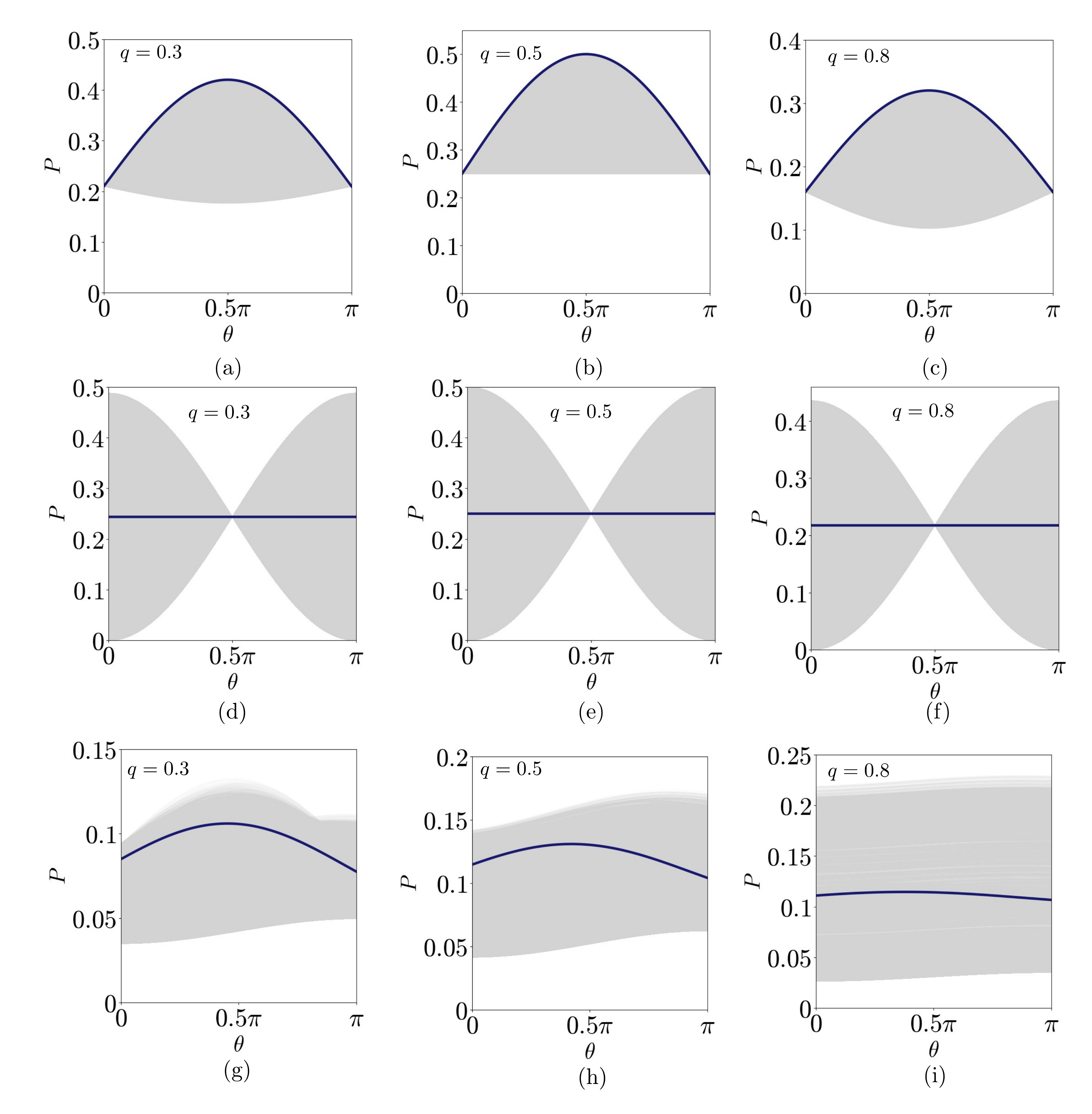}
    \caption{\textbf{Four-qubit code.} $P$ (ordinate) against $\theta$ (abscissa) for different $q$  values when an arbitrary state in the logical subspace is subjected to (a)-(c) bit-flip,  (d)-(f) phase-flip, and (g)-(i) amplitude-damping errors on all qubits, and $t$ is fixed at $t=\pi/2g$. The gray lines correspond to the $H_{SA}$ constituted with randomly sampled $\ket{\Psi_i}\in\mathcal{H}_i$, the details of which are given in Sec.~\ref{subsec:four_qubit}, while the dark continuous lines correspond to the interaction Hamiltonian constituted with the choice of equal superposition of contributing energy eigenstates from $\mathcal{H}_i$ as $\ket{\Psi_i}$ in the cases of (a)-(f), and equal superposition of all energy eigenstates from $\mathcal{H}_i$ as $\ket{\Psi_i}$ in the cases of (g)-(i). The horizontal axes in all figures are in radian.}
    \label{fig:4qubit}
\end{figure*}

Given a specific $\ket{\Psi_{0,0}}$ (i.e., specific values of $\theta$ and $\phi$), and a fixed value of the bit-flip error rate $q$, maximizing  $P$  involves  judiciously choosing $g$, $t$, and $\{b_\alpha\}$. Clearly, from Eq.~(\ref{eq:P_bit_flip}), the optimum time $t=(2\ell+1)\pi/2g$, $\ell=0,1,2,\cdots$, $\forall g$, at which $P=A_1q(1-q)^2+A_2q^2(1-q)$. Also, all of the coefficients $\{b_\alpha\}$ contribute to $P$, and the  choice of optimum set of values of $\{b_\alpha\}$ depends heavily on the values of $\theta$, $\phi$, and $q$. An equal superposition of all \emph{contributing} energy eigenvectors in $\ket{\Psi_1}$, i.e., $b_\alpha=1/\sqrt{6}$ $\forall \alpha$, leads to 
\begin{eqnarray}
    P&=&q(1-q)(1+\sin{\theta}\cos{\phi})/2.
\end{eqnarray}
Our numerical investigation using a sample of $10^5$ randomly generated\footnote{Here, $\ket{\Psi_1}=\sum_{\alpha_1=0}^{5}b_{1,\alpha_1}\ket{E_{1,\alpha_1}}$, where $b_{1,\alpha_1}$ are chosen from a Gaussian distribution of zero mean and unit standard deviation.}  $\ket{\Psi_1}\in\mathcal{H}_1$ reveals that the equal superposition is optimum when $q=0.5$, while it becomes sub-optimal as $q\rightarrow 0$ (i.e., no error), and $q\rightarrow 1$ (i.e., flip definitely occurs on all qubits).  See Fig.~\ref{fig:3qubit}(a)-(c).

\paragraph{Amplitude-damping noise.} The state $\ket{\Psi_{0,0}}$, under the influence of the amplitude damping noise, becomes 
\begin{eqnarray}
    \rho_S^{0}&=&\bigg(\cos^2\frac{\theta}{2}+q^3\sin^2\frac{\theta}{2}\bigg)\ket{000}\bra{000}\nonumber\\&&+(1-q)^3\sin^2\frac{\theta}{2}\ket{111}\bra{111}\nonumber\\
    &&+q(1-q)^2\sin^2\frac{\theta}{2}\big(\ket{011}\bra{011}+\ket{101}\bra{101}\nonumber\\&&+\ket{110}\bra{110}+\ket{100}\bra{100}+\ket{001}\bra{001}\nonumber\\&&
    +\ket{010}\bra{010}\big)+\frac{1}{2}(1-q)^{3/2}\sin\theta\big(\ket{000}\bra{111}\nonumber\\
    &&+\ket{111}\bra{000}\big),
\end{eqnarray}
in the computational basis. Upon evolution under $U$ (see Eq.~(\ref{eq:most_general_situation_closed})) followed by a measurement in the basis $\set{\ket{e_i}:i=0,1}$, we obtain
\begin{eqnarray}
    P&=&\sin^2gt\sin^2\frac{\theta}{2}\Big[q^2(1-q)\sum_{\alpha=0}^2b_\alpha^2 +q(1-q)^2\sum_{\alpha=3}^5b_\alpha^2\Big].\nonumber\\
\end{eqnarray}
The success rates corresponding to a random sampling of $H_{SA}$ via randomly choosing $\ket{\Psi_i}\in\mathcal{H}_1$ are illustrated in Figs.~\ref{fig:3qubit}(d)-(f), where at $q=1/2$, $P$ becomes independent of the choice of $b_\alpha$, as  demonstrated in Fig.~\ref{fig:3qubit}(e). It is important to note that the equal superposition of the contributing energy eigenvectors, given by $b_\alpha=1/\sqrt{6}$ $\forall \alpha$, leads to 
\begin{eqnarray}
    P=\frac{1}{2}q(1-q)\sin^2 gt\sin^2\frac{\theta}{2},
\end{eqnarray}
which are demonstrated in Figs.~\ref{fig:3qubit}(d) and (f).  

\noindent\textbf{Note 1.} Unlike the usual error correction protocols, a non-vanishing success rate of the proposed scheme can always be obtained as long as $q\neq 0,1$. 

\noindent\textbf{Note 2.} Maximum error rate ($q=1$) for all qubits in the code corresponds to the error acting on all three-qubits with certainty. Bit-flip noise with $q=1$ for all qubits in the three-qubit code results in a state in $\mathcal{H}_0$, and thereby corresponding to an uncorrectable error (see Sec.~\ref{sec:Framework}). Similarly, phase flip error on any number of qubits in the three-qubit code does not take the system out of $\mathcal{H}_0$, thereby making the three-qubit code a misfit for correcting phase flip errors via purification.

\subsection{Four-qubit Grassl code} 
\label{subsec:four_qubit}

We now move to the four-qubit Hamiltonian,
\begin{eqnarray}
H_S=-J\left(\sigma^x_1 \sigma^x_2 \sigma^x_3 \sigma^x_4+\sigma^z_3\sigma^z_4+\sigma^z_1\sigma^z_2\right)+3JI,
\end{eqnarray}
with the stabilizers $S_1=\sigma^x_1 \sigma^x_2 \sigma^x_3 \sigma^x_4$, $S_2=\sigma^z_3\sigma^z_4$, and $S_3=\sigma^z_1\sigma^z_2$ corresponding to the four-qubit Grassl code~\cite{nielsen2010,Terhal2015,Roffe2019,Grassl1997}. The structure of the Hilbert space of the $4$-qubit system, in this case, involves $4$ degenerate energy subspaces, including a two-fold degenerate ground-state subspace $\mathcal{H}_0$, inhabited by 
\begin{eqnarray}
     \ket{E_{0,0}}=\frac{1}{\sqrt{2}}(\ket{0000}+\ket{1111}), \nonumber \\
     \ket{E_{0,1}}=\frac{1}{\sqrt{2}}(\ket{0011}+\ket{1100}),
     \label{eq:four_qubit_ground_state_subspace}
\end{eqnarray}
with energy $E_0=0$, and respectively $6$-, $6$-, and $2$-fold degenerate subspaces $\mathcal{H}_1$, $\mathcal{H}_2$, and $\mathcal{H}_3$ with energies $E_1=2$, $E_2=4$, and $E_3=6$. The energy eigenbasis elements spanning these subspaces, which are used for the following calculations, are given in Table~\ref{tab:4_qubit}. Similar to the three-qubit code, we consider $\varrho_S^0=\ket{\Psi_{0,0}}\bra{\Psi_{0,0}}$ (Eq.~(\ref{eq:ground_states})), and use a $4$-level system ($D=3$) as auxiliary.

\paragraph{Bit-flip error.} Unlike the three-qubit code, in the present case, application of a subset of possible two-qubit errors, constituted of $\sigma^x_1\sigma^x_2$, $\sigma^x_3\sigma^x_4$ as well as error on all four qubits, i.e., $\sigma^x_1\sigma^x_2\sigma^x_3\sigma^x_4$ keeps $\ket{\Psi_{0,0}}$ within $\mathcal{H}_0$, thereby escaping the purification process. Assuming that error applies to all qubits in the code and  following the same line of calculations as shown in the case of three-qubit code,  we obtain 
\begin{eqnarray}
P&=&(q-3q^2+4q^3-2q^4)\sin^2 g_1t\sum_{\alpha=1}^2W(b_{1,\alpha},b_{1,\alpha+2})\nonumber\\&&+2q^2(1-q)^2W(b_{2,3},b_{2,4})\sin^2 g_2t,
\label{eq:general_P_four_qubit}
\end{eqnarray}
where we have defined $W(x,y)=x^2+2xy\sin{\theta}\cos{\phi}+y^2$,
and have used the energy eigenbasis specified in Eq.~(\ref{eq:four_qubit_ground_state_subspace}) and Table~\ref{tab:4_qubit} for the calculations, and have used $\{\ket{e_i},i=0,1,2,3\}$ as the measurement basis.   

Note that in contrast to the case of the three-qubit code, Eq.~(\ref{eq:general_P_four_qubit}) indicates that only a subset of the energy-eigenstates $\ket{E_{i,\alpha}}\in\mathcal{H}_i$ contribute to $P$. Assuming equal superposition of the contributing energy-eigenvectors as  $\ket{\Psi_i}\in\mathcal{H}_i$, i.e, $b_{1,1}=b_{1,2}=b_{1,3}=b_{1,4}=1/2$, and $b_{2,3}=b_{2,4}=1/\sqrt{2}$, one obtains 
\begin{eqnarray}
    P&=&(1+\sin\theta \cos \phi)\Big[ (q-3q^2+4q^3-2q^4)\sin^2g_1t\nonumber \\
&&+2q^2(1-q)^2\sin^2g_2t \Big].
\end{eqnarray}
which, for $g_1=g_2=g$, results in 
\begin{eqnarray}
    P&=& q(1-q)\sin^2gt(1+\sin\theta \cos \phi).
\end{eqnarray}
It is worthwhile to note that the use of an equal superposition of all energy eigenstates from all energy subspaces (i.e., $b_{i,\alpha}=1/\sqrt{d_i}$, $\forall i=1,2,3$) provides   
$P^\prime=\sin^2 gt (1+\sin\theta \cos \phi)(q-2q^2+2q^3-q^4)/3$, leading to a lesser success rate than $P$ when $t=(2\ell+1)\pi/2g$ ($\ell=0,1,2,\cdots$), and for fixed $(\theta,\phi)$. We carry out a numerical investigation over a sample of $2\times 10^4$ random states  $\ket{\Psi_1}\in\mathcal{H}_1$ and $\ket{\Psi_2}\in\mathcal{H}_2$, i.e., the contributing energy eigenspaces,  by randomly generating $10^4$ states of the form $\ket{\Psi_1}=\sum_{\alpha_1=1}^{4}b_{1,\alpha_1}\ket{E_{1,\alpha_1}}$ while choosing $\ket{\Psi_2}=\left(\ket{E_{2,3}}+\ket{E_{2,4}}\right)/\sqrt{2}$, and  $10^4$ states of the form $\ket{\Psi_2}=b_{2,3}\ket{E_{2,3}}+b_{2,4}\ket{E_{2,4}}$ while choosing $\ket{\Psi_1}=2^{-1}\sum_{\alpha_1=1}^4\ket{E_{1,\alpha_1}}$. 
We observe that for all $q$, choice of the equal superposition of the \emph{contributing} energy eigenvectors as $\ket{\Psi_i}$ provides the best success rate of the protocol (see Fig.~\ref{fig:4qubit}(a)-(c)). 

\paragraph{Phase-flip error.} Similar to the BF noise, The success rate corresponding to the PF noise on the logical subspace of the four-qubit code can be calculated using the same energy eigenbasis given in Table~\ref{tab:4_qubit}, and is given by
\begin{eqnarray}
    P&=&4\left(q-3q^2+4q^3-2q^4\right)\sin^2g_1t \nonumber \\
    &&\times(b_{10}^2\cos^2\frac{\theta}{2}+b_{15}^2\sin^2\frac{\theta}{2}),
    \label{eq:4_qubit_PF_success_rate}
\end{eqnarray}
where only two of the energy eigenstates of $H_S$, namely, $\ket{E_{1,0}}$ and $\ket{E_{1,5}}$, corresponding to energy higher than $0$ contribute to $P$. Considering $\ket{\Psi_1}=\left(\ket{E_{1,0}}+\ket{E_{1,5}}\right)/\sqrt{2}$, the success rate becomes 
\begin{eqnarray}
    P&=&4\left(q-3q^2+4q^3-2q^4\right)\sin^2g_1t,
\end{eqnarray}
while it is clear from Eq.~(\ref{eq:4_qubit_PF_success_rate}) that for a fixed $q$, the maximum $P$ for a fixed $\theta$ occurs for either $b_{10}=0$, or $b_{15}=0$, leading to the maximum success rate given by
\small 
\begin{eqnarray}
    P&=&\left\{\begin{array}{ll}
       4\left(q-3q^2+4q^3-2q^4\right)\sin^2g_1t \cos^2\frac{\theta}{2}  &\text{for }\theta\leq\frac{\pi}{2},  \\
       4\left(q-3q^2+4q^3-2q^4\right)\sin^2g_1t \sin^2\frac{\theta}{2}  & \text{for }\theta>\frac{\pi}{2}. 
    \end{array}
    \right.\nonumber\\
\end{eqnarray}\normalsize 
Our numerical investigation with a sample of $10^5$ randomly generated states of the form $\ket{\Psi_1}=b_{1,0}\ket{E_{1,0}}+b_{1,5}\ket{E_{1,5}}$ indicates that the choice of equal superposition of the contributing energy eigenstates as $\ket{\Psi_1}$ is sub-optimal for purifyng the logical state with a high success rate. See Fig.~\ref{fig:3qubit}(d)-(f).

\paragraph{Amplitude-damping noise.} The success rate $P$ corresponding to the four-qubit Grassl code subjected to the amplitude-damping noise could be derived in a similar fashion, although the expressions are cumbersome and we refrain from including them in the text. In Fig.~\ref{fig:4qubit}(d)-(f), we plot $P$ as functions of $\theta$ for different values of $q$, where the archetype states  are chosen randomly. Note that without the analytical insight, it is difficult to extract information regarding the contributing energy eigenstates from each subspace. In this situation, we carry out a numerical analysis with a total of $3\times10^4$ random states, where during sampling $10^4$ random states of the form, say, $\ket{\Psi_i}\in\mathcal{H}_i$, $\ket{\Psi_j}\in\mathcal{H}_j$ ($j\neq i$)  is fixed at $\ket{\Psi_j}=d_j^{-1/2}\sum_{\alpha_j=0}^{d_j}\ket{E_{j,\alpha_j}}$.   
It is evident from Fig.~\ref{fig:4qubit} that the choice of the equal superposition of all energy eigenvectors $\ket{E_{i,\alpha}}$ spanning $\mathcal{H}_i$ for all $i=1,2,3$ serves as a suboptimal choice.

\begin{figure*}
    \centering
    \includegraphics[width=0.8\linewidth]{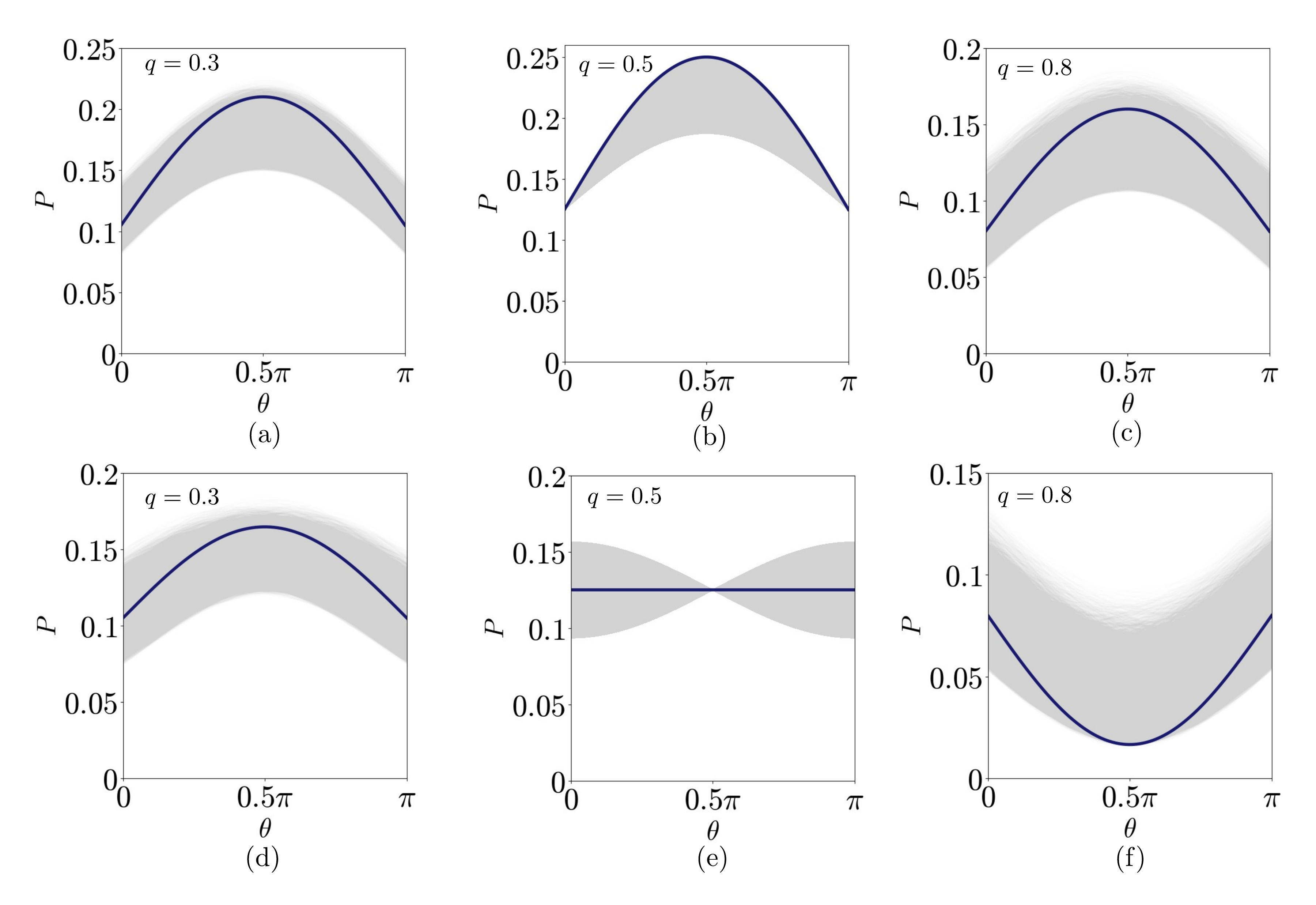}
    \caption{\textbf{Five-qubit code.} $P$ (ordinate) vs  $\theta$ (abscissa) for different error rates $q$.
    (a)-(c) Bit-flip, and  (d)-(f) phase-flip errors on all qubits. 
    All other specifications are the same as in Fig. \ref{fig:4qubit}. The horizontal axes in all figures are in radian, and $t=\pi/2g$ for all figures.}
    \label{fig:5qubit}
\end{figure*}


\subsection{Five-qubit perfect code} 
\label{subsec:five_qubit}

The  Hamiltonian corresponding to the five-qubit code~\cite{nielsen2010,Terhal2015,Roffe2019,Laflamme1996}, given by
\begin{eqnarray}
    H_S&=&-J(\sigma^x_1\sigma^z_2\sigma^z_3\sigma^x_4+\sigma^x_2\sigma^z_3\sigma^z_4\sigma^x_5+\sigma^x_1\sigma^x_3\sigma^z_4\sigma^z_5\nonumber \\&&+\sigma^z_1\sigma^x_2\sigma^x_4\sigma^z_5)+4J,
\end{eqnarray}
involves the stabilizers $S_1=\sigma^x_1\sigma^z_2\sigma^z_3\sigma^x_4$, $S_2=\sigma^x_2\sigma^z_3\sigma^z_4\sigma^x_5$, $\sigma^x_1\sigma^x_3\sigma^z_4\sigma^z_5$, and $S_4=\sigma^z_1\sigma^x_2\sigma^x_4\sigma^z_5$. The Hilbert space corresponding to $H_S$ can be divided into $5$ degenerate energy subspaces (i.e., $D=4$), among which the ground state subspace $\mathcal{H}_0$ is two fold degenerate with $E_0=0$, and is spanned by $\ket{E_{0,0}}$ and $\ket{E_{0,1}}$ given in the computational basis by
\begin{eqnarray}
   \ket{E_{0,0}}&=&-\frac{1}{4}\Bigg[\ket{00001}+\ket{00010}+\ket{00100}+\ket{00111}\nonumber\\&&+\ket{01000}-\ket{01011}-\ket{01101}+\ket{01110}\nonumber\\&&+\ket{10000}+\ket{10011}-\ket{10101}-\ket{10110}\nonumber\\&&+\ket{11001}-\ket{11010}+\ket{11100}-\ket{11111}\Bigg], \nonumber \\
   \ket{E_{0,1}}&=&\left[\bigotimes_{j=1}^5\sigma^x_j\right]\ket{E_{0,0}},
\end{eqnarray}
The rest of the subspaces $\mathcal{H}_{i}$, $i=1,2,3,4$, are respectively $8$-, $12$-, $8$, and $2$-fold degenerate, with energies $E_1=2J$, $E_2=4J$, $E_3=6J$, and $E_4=8J$ respectively.  

\begin{table*}[]
    \centering
    \begin{tabular}{|l|l|l|l|l|}
    \hline
    Subspaces & $\mathcal{H}_1$ & $\mathcal{H}_2$ & $\mathcal{H}_3$ & $\mathcal{H}_4$ \\
    \hline 
    Spanning elements &
    \begin{tabular}{l}
    $\ket{E_{1,0}}=\sigma^x_1\ket{E_{0,0}}$\\
    $\ket{E_{1,1}}=\sigma^x_2\ket{E_{0,0}}$\\
    $\ket{E_{1,2}}=\sigma^x_1\ket{E_{0,1}}$\\
    $\ket{E_{1,3}}=\sigma^x_2\ket{E_{0,1}}$\\
    $\ket{E_{1,4}}=\sigma^x_2\sigma^x_3\ket{E_{0,0}}$\\
    $\ket{E_{1,5}}=\sigma^x_1\sigma^x_5\ket{E_{0,0}}$\\
    $\ket{E_{1,6}}=\sigma^x_2\sigma^x_3\ket{E_{0,1}}$\\
    $\ket{E_{1,7}}=\sigma^x_1\sigma^x_5\ket{E_{0,1}}$
    \end{tabular}     &  
    \begin{tabular}{l}
    $\ket{E_{2,0}}=\sigma^x_3\ket{E_{0,0}}$\\
    $\ket{E_{2,1}}=\sigma^x_4\ket{E_{0,0}}$\\
    $\ket{E_{2,2}}=\sigma^x_5\ket{E_{0,0}}$\\
    $\ket{E_{2,3}}=\sigma^x_3\ket{E_{0,1}}$\\
    $\ket{E_{2,4}}=\sigma^x_4\ket{E_{0,1}}$\\
    $\ket{E_{2,5}}=\sigma^x_5\ket{E_{0,1}}$\\
    $\ket{E_{2,6}}=\sigma^x_1\sigma^x_2\ket{E_{0,0}}$\\
    $\ket{E_{2,7}}=\sigma^x_3\sigma^x_4\ket{E_{0,0}}$\\
    $\ket{E_{2,8}}=\sigma^x_4\sigma^x_5\ket{E_{0,0}}$\\
    $\ket{E_{2,9}}=\sigma^x_1\sigma^x_2\ket{E_{0,1}}$\\
    $\ket{E_{2,10}}=\sigma^x_3\sigma^x_4\ket{E_{0,1}}$\\
    $\ket{E_{2,11}}=\sigma^x_4\sigma^x_5\ket{E_{0,1}}$
    \end{tabular}     &
    \begin{tabular}{l}
    $\ket{E_{3,0}}=\sigma^x_1\sigma^x_3\ket{E_{0,0}}$\\
    $\ket{E_{3,1}}=\sigma^x_1\sigma^x_4\ket{E_{0,0}}$\\
    $\ket{E_{3,2}}=\sigma^x_2\sigma^x_4\ket{E_{0,0}}$\\
    $\ket{E_{3,3}}=\sigma^x_2\sigma^x_5\ket{E_{0,0}}$\\
    $\ket{E_{3,4}}=\sigma^x_1\sigma^x_3\ket{E_{0,1}}$\\
    $\ket{E_{3,5}}=\sigma^x_1\sigma^x_4\ket{E_{0,1}}$\\
    $\ket{E_{3,6}}=\sigma^x_2\sigma^x_4\ket{E_{0,1}}$\\
    $\ket{E_{3,7}}=\sigma^x_2\sigma^x_5\ket{E_{0,1}}$
    \end{tabular}     &
    \begin{tabular}{l}
    $\ket{E_{4,0}}=\sigma^x_3\sigma^x_5\ket{E_{0,0}}$\\
    $\ket{E_{4,1}}=\sigma^x_3\sigma^x_5\ket{E_{0,1}}$
    \end{tabular}  \\
    \hline 
    \end{tabular}
    \caption{Elements of the energy eigenbasis of the five-qubit code that span the different subspaces $\mathcal{H}_i$, $i=1,2,3,4$, used in the calculation of success rate corresponding to the bit-flip noise.}
    \label{tab:5_qubit_BF}
\end{table*}

 \begin{table*}[]
    \centering
    \begin{tabular}{|l|l|l|l|l|}
    \hline
    Subspaces & $\mathcal{H}_1$ & $\mathcal{H}_2$ & $\mathcal{H}_3$ & $\mathcal{H}_4$ \\
    \hline 
    Spanning elements &
    \begin{tabular}{l}
    $\ket{E_{1,0}}=\sigma^z_3\ket{E_{0,0}}$\\
    $\ket{E_{1,1}}=\sigma^z_5\ket{E_{0,0}}$\\
    $\ket{E_{1,2}}=\sigma^z_3\ket{E_{0,1}}$\\
    $\ket{E_{1,3}}=\sigma^z_5\ket{E_{0,1}}$\\
    $\ket{E_{1,4}}=\sigma^z_1\sigma^z_3\ket{E_{0,0}}$\\
    $\ket{E_{1,5}}=\sigma^z_2\sigma^z_5\ket{E_{0,0}}$\\
    $\ket{E_{1,6}}=\sigma^z_1\sigma^z_3\ket{E_{0,1}}$\\
    $\ket{E_{1,7}}=\sigma^z_2\sigma^z_5\ket{E_{0,1}}$
    \end{tabular}     &  
    \begin{tabular}{l}
    $\ket{E_{2,0}}=\sigma^z_1\ket{E_{0,0}}$\\
    $\ket{E_{2,1}}=\sigma^z_2\ket{E_{0,0}}$\\
    $\ket{E_{2,2}}=\sigma^z_4\ket{E_{0,0}}$\\
    $\ket{E_{2,3}}=\sigma^z_1\ket{E_{0,1}}$\\
    $\ket{E_{2,4}}=\sigma^z_2\ket{E_{0,1}}$\\
    $\ket{E_{2,5}}=\sigma^z_4\ket{E_{0,1}}$\\
    $\ket{E_{2,6}}=\sigma^z_1\sigma^1_4\ket{E_{0,0}}$\\
    $\ket{E_{2,7}}=\sigma^z_2\sigma^z_4\ket{E_{0,0}}$\\
    $\ket{E_{2,8}}=\sigma^z_3\sigma^z_5\ket{E_{0,0}}$\\
    $\ket{E_{2,9}}=\sigma^z_1\sigma^z_4\ket{E_{0,1}}$\\
    $\ket{E_{2,10}}=\sigma^z_3\sigma^z_4\ket{E_{0,1}}$\\
    $\ket{E_{2,11}}=\sigma^z_3\sigma^z_5\ket{E_{0,1}}$
    \end{tabular}     &
    \begin{tabular}{l}
    $\ket{E_{3,0}}=\sigma^z_1\sigma^z_5\ket{E_{0,0}}$\\
    $\ket{E_{3,1}}=\sigma^z_2\sigma^z_3\ket{E_{0,0}}$\\
    $\ket{E_{3,2}}=\sigma^z_3\sigma^z_4\ket{E_{0,0}}$\\
    $\ket{E_{3,3}}=\sigma^z_4\sigma^z_5\ket{E_{0,0}}$\\
    $\ket{E_{3,4}}=\sigma^z_1\sigma^z_5\ket{E_{0,1}}$\\
    $\ket{E_{3,5}}=\sigma^z_2\sigma^z_3\ket{E_{0,1}}$\\
    $\ket{E_{3,6}}=\sigma^z_3\sigma^z_4\ket{E_{0,1}}$\\
    $\ket{E_{3,7}}=\sigma^z_4\sigma^z_5\ket{E_{0,1}}$
    \end{tabular}     &
    \begin{tabular}{l}
    $\ket{E_{4,0}}=\sigma^z_1\sigma^z_2\ket{E_{0,0}}$\\
    $\ket{E_{4,1}}=\sigma^z_1\sigma^z_2\ket{E_{0,1}}$
    \end{tabular}  \\
    \hline 
    \end{tabular}
    \caption{Elements of the energy eigenbasis of the five-qubit code that span the different subspaces $\mathcal{H}_i$, $i=1,2,3,4$, used in the calculation of success rate corresponding to the phase-flip noise.}
    \label{tab:5_qubit_PF}
\end{table*}

\paragraph{Bit-flip error.} The structure of the Hilbert space demands
a qudit of dimension $5$ ($D+1=5$) as the auxiliary. While we refrain from including $\rho_{SA}(t)$ to keep the text uncluttered, it is worth mentioning that similar to the case of the three-qubit code, here also the success rate receives contribution from all diagonal terms of $\rho_{SA}(t)$ except the ones originating from the single-qubit bit-flip error on no, or on all qubits, as they eventually transform $\ket{\Psi_{0,0}}$ to states belonging to $\mathcal{H}_0$. Choosing $g_i=g$ $\forall i$, the success probability can be calculated as
\begin{eqnarray}
   P&=&A_1q(1-q)^4+B_1q^2(1-q)^3+B_2q^3(1-q)^2\nonumber\\&&+A_2q^4(1-q),
\end{eqnarray}
with 
\begin{widetext}
\begin{eqnarray}    A_1&=&\sin^2g_1t\sum_{\alpha=0}^1U_+(b_{1,\alpha},b_{1,\alpha+2})
+\sin^2g_2t\sum_{\alpha=0}^2U_+(b_{2,\alpha},b_{2,\alpha+3}),
\nonumber\\
A_2&=&\sin^2g_1t\sum_{\alpha=0}^1V(b_{1,\alpha},b_{1,\alpha+2})+\sin^2g_2t\sum_{\alpha=0}^2V(b_{2,\alpha},b_{2,\alpha+3}),
\end{eqnarray}
and
\begin{eqnarray}
B_1&=&\sin^2g_1t\sum_{\alpha=4}^5U_+(b_{1,\alpha},b_{1,\alpha+2})+\sin^2g_2t\sum_{\alpha=6}^8U_+(b_{2,\alpha},b_{2,\alpha+3})+\sin^2g_3t\sum_{\alpha=0}^3U_+(b_{3,\alpha},b_{3,\alpha+4})\nonumber\\&&+\sin^2g_4tU_+(b_{4,0},b_{4,1}),\nonumber\\
B_2&=&\sin^2g_1t\sum_{\alpha=4}^5V(b_{1,\alpha},b_{1,\alpha+2})+\sin^2g_2t\sum_{\alpha=6}^8V(b_{2,\alpha},b_{2,\alpha+3})+\sin^2g_3t\sum_{\alpha=0}^3V(b_{3,\alpha},b_{3,\alpha+4})\nonumber\\&&+\sin^2g_4tV(b_{4,0},b_{4,1}),
\end{eqnarray}
\end{widetext}
where $U_+$ and $V$ are as defined in Eq.~(\ref{eq:UV_definitions}), and the spanning energy eigenbasis elements corresponding to $\mathcal{H}_i$, $i=1,2,3,4$, used in this calculation are as given in Table~\ref{tab:5_qubit_BF}. 

\begin{figure*}
    \centering
    \includegraphics[width=0.6\linewidth]{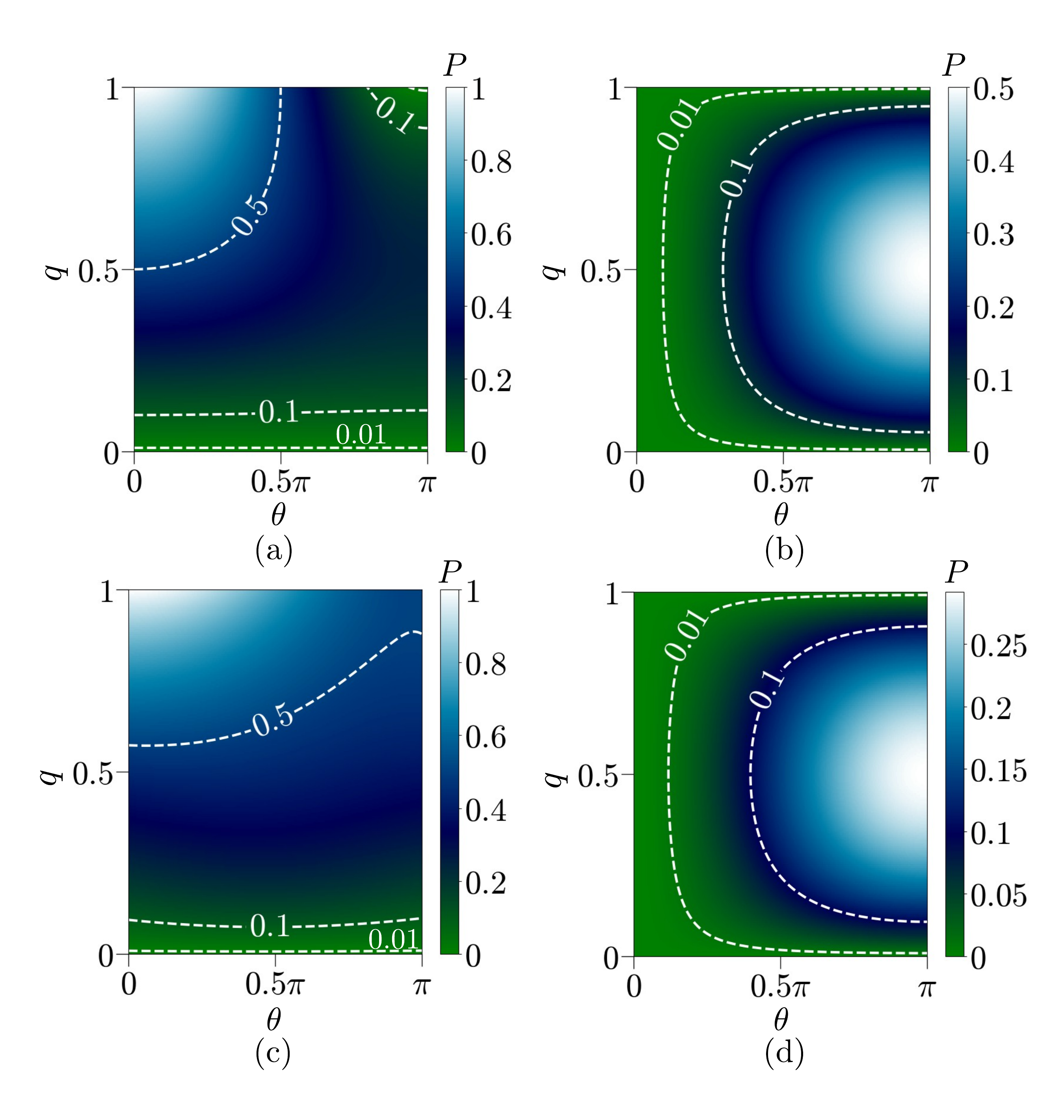}
    \caption{\textbf{1D isotropic Heisenberg model.} 
     Map plot of  $P$ with  $\theta$ (abscissa)  and $q$ (ordinate) in the cases of (a)-(b) two-qubit isotropic Heisenberg encoding under (a) bit-flip and (b) phase-flip noise, and (c)-(d) two-qubit isotropic Heisenberg encoding under (c) bit-flip and (d) phase-flip noise, with $t=\pi/2g$ for all figures. The dashed lines mark boundaries of the regions where $P>0.01$, $0.1$, and $0.5$.  The horizontal axis is in radian.}
    \label{fig:Heisenberg_bf_pf}
\end{figure*}

For $g_i=g, \forall i=1,2,3,4$, and choosing $\ket{\Psi_i} \in\mathcal{H}_i$ as states with equal superpositions of contributing energy eigenstates, i.e., $b_{1,\alpha}=1/\sqrt{8}$, $b_{2,\alpha}=1/\sqrt{12}$, $b_{3,\alpha}=1/\sqrt{8}$ and $b_{4,\alpha}=1/\sqrt{2}$ $\forall \alpha$, the success rate is given by 
\begin{eqnarray}
    P&=&\frac{1}{2}\sin^2gt(1+\sin\theta\cos \phi)\big[q(1-q)^4 \nonumber \\
    &&+q^4(1-q)+3q^2(1-q)^3+3q^3(1-q)^2\big]. 
\end{eqnarray}
We perform numerical investigation of the dependence of $P$ on the choice of $\ket{\Psi_i}$ by randomly sampling $10^4$ states $\ket{\Psi_i}\in\mathcal{H}_i$,for each $i=1,2,3,4$, while $\ket{\Psi_j}\in\mathcal{H}_j$ for all $j\neq i$ is set to be equal superposition of all $\ket{E_{j,\alpha_j}}\in\mathcal{H}_j$.  Our results indicate that the choice of the equal superposition of the contributing energy eigenstates for constructing $H_{SA}$ is sub-optimal for five-qubit code as well (see Fig.~\ref{fig:5qubit}(a)-(c)).

\paragraph{Phase-flip noise.} On the other hand, simplification of the calculation applying PF noise of strength $q$ on all qubits demands a different set of energy eigenvectors $\ket{E_{i,\alpha}}$ spanning $\mathcal{H}_i$ ($i=1,2,3,4$), which is given in Table~\ref{tab:5_qubit_PF}. In this case, the success rate reads as
\begin{eqnarray}
   P&=&A_1q(1-q)^4+B_1q^2(1-q)^3+B_2q^3(1-q)^2\nonumber\\&&+A_2q^4(1-q),
\end{eqnarray}
with 
\begin{widetext}
\begin{eqnarray}    A_1&=&\sin^2g_1t\sum_{\alpha=0}^1U_+(b_{1,\alpha},b_{1,\alpha+2})
+\sin^2g_2t\sum_{\alpha=0}^2U_+(b_{2,\alpha},b_{2,\alpha+3}),
\nonumber\\
A_2&=&\sin^2g_1t\sum_{\alpha=0}^1U_-(b_{1,\alpha},b_{1,\alpha+2})+\sin^2g_2t\sum_{\alpha=0}^2U_{-}(b_{2,\alpha},b_{2,\alpha+3}),
\end{eqnarray}
and
\begin{eqnarray}
B_1&=&\sin^2g_1t\sum_{\alpha=4}^5U_+(b_{1,\alpha},b_{1,\alpha+2})+\sin^2g_2t\sum_{\alpha=6}^8U_+(b_{2,\alpha},b_{2,\alpha+3})+\sin^2g_3t\sum_{\alpha=0}^3U_+(b_{3,\alpha},b_{3,\alpha+4})\nonumber\\&&+\sin^2g_4tU_+(b_{4,0},b_{4,1}),\nonumber\\
B_2&=&\sin^2g_1t\sum_{\alpha=4}^5U_-(b_{1,\alpha},b_{1,\alpha+2})+\sin^2g_2t\sum_{\alpha=6}^8U_-(b_{2,\alpha},b_{2,\alpha+3})+\sin^2g_3t\sum_{\alpha=0}^3U_-(b_{3,\alpha},b_{3,\alpha+4})\nonumber\\&&+\sin^2g_4tU_-(b_{4,0},b_{4,1}),
\end{eqnarray}
\end{widetext}
where $U_{\pm}$ is as defined in Eq.~(\ref{eq:UV_definitions}). Setting $g_i=g \forall i=1,2,3,4$, and choosing $\ket{\Psi_i} \in\mathcal{H}_i$ as states with equal superpositions of contributing energy eigenstates, i.e., $b_{1,\alpha}=1/\sqrt{8}$, $b_{2,\alpha}=1/\sqrt{12}$, $b_{3,\alpha}=1/\sqrt{8}$ and $b_{4,\alpha}=1/\sqrt{2}$ $\forall \alpha$, the success rate is given by 
\begin{eqnarray}
    P&=&\frac{1}{2}\sin^2gt \Big[(1+\sin\theta\cos \phi)\big(q(1-q)^4 +3q^2(1-q)^3\big)\nonumber \\
    &&+(1-\sin \theta \cos \phi)\big(q^4(1-q)+3q^3(1-q)^2\big)\Big]. 
\end{eqnarray}
Our numerical investigation of the dependence of $P$ on the choice of $\ket{\Psi_i}$ (where the details corresponding to the numerical invetigation remains the same as the case of the BF noise) indicates that the choice of the equal superposition of the contributing energy eigenstates for constructing $H_{SA}$ is sub-optimal for five-qubit codes (see Fig.~\ref{fig:5qubit}(d)-(f)).


\section{QEC in one-dimensional Isotropic Heisenberg model}
\label{sec:heisenberg_model}

The $N$-qubit 1D isotropic Heisenberg Hamiltonian in a magnetic field along the $z$ direction reads~\cite{giamarchi2004,Franchini2017}
\begin{eqnarray}
\label{eq:system_Hamiltonian_one_rung}
H=\frac{1}{4}\sum_{j=1}^{N}\vec{\sigma}_{j}.\vec{\sigma}_{j+1}-\frac{h}{2}\sum_{j=1}^N\sigma_{j}^z-E_g,
\end{eqnarray}
while the constant $E_g$ ensures that the ground state energy $E_0$ of $H$ is zero. For arbitrary $N$ with large $h$, the ground state of the system is fully polarized, given by $\ket{E_0}=\ket{0}^{\otimes N}$, with $E_g=N\left(\frac{1}{4}-\frac{h}{2}\right)$. For $N>2$ and even, the ground state of the system for $h<h_0=2$ ($h<h_0=1$ for $N=2$) is given by 
\begin{eqnarray}
\ket{E_1}=\frac{1}{\sqrt{N}}\sum_{j=1}^{N}(-1)^{j}\ket{j},  
\label{eq:all_up_except_one}
\end{eqnarray}
where $\ket{j}=\sigma^x_{j} \ket{0}^{\otimes N}$. At $h=h_0$,  the ground state subspace is doubly degenerate, which has been used as a logical qubit in the case of quantum state transfer through 2D rectangular lattices. While we focus only on isotropic Heisenberg model with even $N$ at $h=h_0$ in this paper, it is worthwhile to note that our formalism is applicable to the case of odd $N$ also, in which case the degeneracy structure of the Hilbert space is different, depending on the value of $h$.

For even $N$,  the doubly degenerate ground state subspace $\mathcal{H}_0$ at $h=h_0$ is spanned by  $\ket{E_{0,0}}(=\ket{E_0})$ and $\ket{E_{0,1}}(=\ket{E_1})$, according to the notations introduced in Sec.~\ref{sec:Framework}.  We now demonstrate the applicability of our protocol in the specific case of $N=2$ ($h_0=1$), for which the ground state subspace $\mathcal{H}_0$ corresponding to the ground state energy $E_0=0$ (when $E_g=3/4$) is spanned by \begin{eqnarray}
    \ket{E_{0,0}}&=&\ket{00}, \ket{E_{0,1}}=\frac{1}{\sqrt{2}}(\ket{01}-\ket{10}).
\end{eqnarray}   
The excited states of the two-qubit isotropic Heisenberg model at $h=1$ are non-degenerate, given by 
\begin{eqnarray}
    \ket{E_{1,0}}&=&\frac{1}{\sqrt{2}}(\ket{01}+\ket{10}),\ket{E_{2,0}}=\ket{11}),
\end{eqnarray}
corresponding to the energies $E_1=1$ and $E_2=2$, respectively, implying the requirement of an auxiliary qutrit for the protocol. We specifically consider the state $\ket{\Psi_{0,0}}$ (see Eq.~(\ref{eq:ground_states})) as the resource state, which is subjected to noise, and construct $H_{SA}$ assuming $\ket{\Psi_i}=\ket{E_{i,0}}$, $i=1,2$ (see Eq.~(\ref{eq:archetype_states})).

\begin{figure*}
    \centering
    \includegraphics[width=0.9\linewidth]{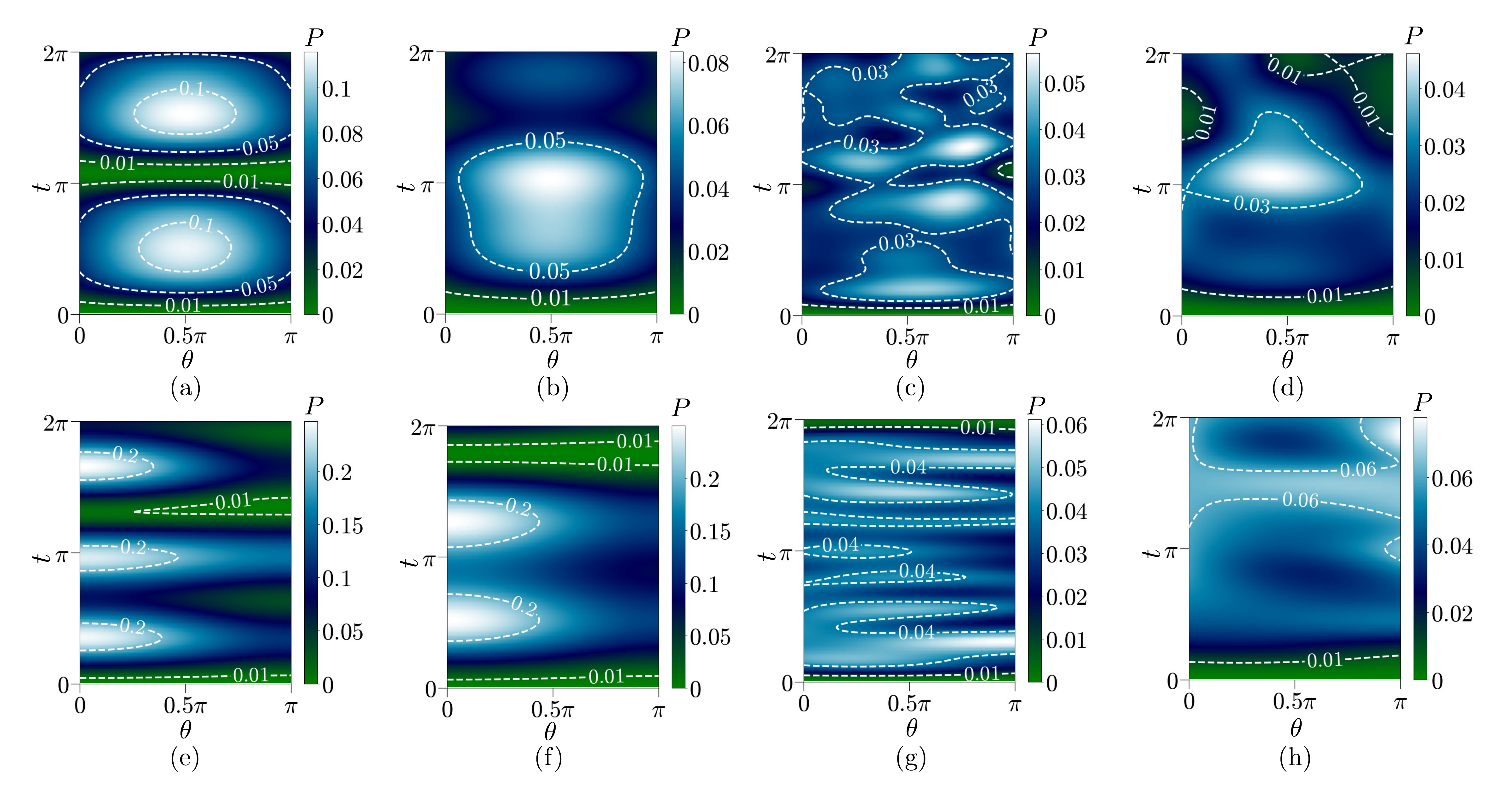}
    \caption{\textbf{Purification with an auxiliary qubit.} Landscapes of $P$ with respect to  $\theta$ (horizontal axis) and $t$ (vertical axis) in the case of the four-qubit code with (a) Prescription 1 and (b) Prescription 2, and the five-qubit code with (c) Prescription 1 and (d) Prescription 2 under the bit-flip noise of strength $q=0.5$ on all qubits.  The same for (e)-(f) the two-qubit isotropic Heisenberg encoding, and (g)-(h) the four-qubit isotropic Heisenberg encoding with (e)-(g) Prescription 1 and (f)-(h) Prescription 2 are also shown. The vertical axis is in the unit of $g^{-1}$, while the horizontal axis is in radian.}
    \label{fig:qubit_auxiliary}
\end{figure*}

The success rate, in this case, turns out to be 
\begin{eqnarray}
    P&=&q(1-q)\cos^2\frac{\theta}{2}\sin^2g_1t+q\bigg[\sin^2\frac{\theta}{2}\nonumber\\&&+q\cos \theta\bigg]\sin^2g_2t,
\end{eqnarray}
corresponding to the bit-flip noise, while for the phase-flip noise, 
\begin{eqnarray}
    P=2q(1-q)\sin^2\frac{\theta}{2}\sin^2g_1t. 
\end{eqnarray}
However,  in contrast, the two-qubit Heisenberg encoding fails to correct amplitude-damping noise as the error does not take the resource state outside the logical subspace. This is evident from the applications of the corresponding Krauss operators on $\ket{\Psi_{0,0}}$, as given below.  
\begin{eqnarray}
    K_0K_0\ket{\Psi_{0,0}}&=&\cos\frac{\theta}{2}\ket{E_{0,0}}+\sqrt{1-\gamma}\text{e}^{i\phi}\sin \frac{\theta}{2}\ket{E_{0,1}}, \nonumber \\
    K_1K_0\ket{\Psi_{0,0}}&=&-\sqrt{\frac{\gamma}{2}}\text{e}^{i\phi}\sin \frac{\theta}{2}\ket{E_{0,0}}, \nonumber \\
    K_0K_1\ket{\Psi_{0,0}}&=&\sqrt{\frac{\gamma}{2}}\text{e}^{i\phi}\sin \frac{\theta}{2}\ket{E_{0,0}}, \nonumber \\
    K_1K_1\ket{\Psi_{0,0}}&=&0.
\end{eqnarray}

With increasing $N$, the calculation is cumbersome for the 1D isotropic Heisenberg model as the value of $D$ is large due to scarcity of degenerate excited state subspaces at $h=h_0$. For example, with $N=4$ and $E_g=3$, the Hilbert space is divided into a total of $8$ energy subspaces, including $\mathcal{H}_0$ corresponding to $E_0=0$, spanned by
\begin{eqnarray}
    \ket{E_{0,0}}&=&\ket{0000},\text{ and } \nonumber \\
    \ket{E_{0,1}}&=&\frac{1}{2}(\ket{0001}-\ket{0010}+\ket{0100}-\ket{1000}),
\end{eqnarray}
while the excited state subspaces $\mathcal{H}_i$s  ($i=1,2,\cdots,5$) being respectively $3$, $2$, $3$, $2$, and $2$-fold degenerate, and $\mathcal{H}_{6}$ and $\mathcal{H}_7$ being non-degenerate, with $E_i=i$, $i=1,2,\cdots,6$, and $E_7=8$ (see~\cite{Pal_2011} for the full set of energy eigenbasis). Fig.~\ref{fig:Heisenberg_bf_pf} depicts the variations of $P$ with $\theta$ in the case of the two- and four-qubit encoding with the 1D isotropic Heisenberg model under the BF and the PF noises, where in the case of the four-qubit encoding, $b_{i,\alpha}$ is taken to be $1\sqrt{d_i}$ in the chosen basis, while constructing $\ket{\Psi_i}$ corresponding to $\mathcal{H}_i$ for which $d_i>1$. 

Fig.~\ref{fig:Heisenberg_bf_pf}(a)-(d) depicts the variations of $P$ as functions of $\theta$ and $q$ in the case of the two- and four-qubit isotropic Heisenberg encoding under the bit- and phase-flip noises. These figures indicate that the isotropic Heisenberg encoding is unable to correct BF noise if the noise strength is low. On the other hand, for the phase-flip noise, $\theta=0$ is uncorrectable as $\ket{0}^{\otimes N}$ remains in the logical subspace under phase-flip error on any of the qubits. In general, the success rate decreases with an increase in $N$ in the case of the phase-flip noise, while in contrast, such trend is absent in the case of the bit-flip noise.

\section{Reduction of resource in QEC: Purification with auxiliary qubit}
\label{sec:purification_with_qubit}

\emph{Can purification of arbitrary mixed states of a quantum system into its ground state be achieved using less resource by employing an auxiliary qubit instead of a qudit?} Let us assume that the qubit Hamiltonian is given by 
\begin{eqnarray}
    H_A=E^\prime\ket{1}\bra{1},
\end{eqnarray}
where $E^\prime$ is the energy of the qubit. The reduction in the available number of levels require connecting transitions corresponding to all energy subspaces with the auxiliary qubit, resulting in a reconstruction of the system-auxiliary interaction. In this paper, we consider two specific proposals for reconstructions. We carry out the investigation (a) for the four- and five-qubit quantum codes, and (b) for the two- and four-qubit isotropic Heisenberg model under the influence of the BF noise, and answer the above question affirmatively. 

\paragraph{Prescription 1.} In the first proposal, the system-auxiliary interaction Hamiltonian is given by  
\begin{eqnarray}
    H_{SA}&=& \sum_{i=1}^Dg_i\left(\ket{\Psi_{0,\gamma}e_1}\bra{\Psi_ie_0}+\text{h.c.}\right),
    \label{eq:system_auxiliary_qubit_interaction}
\end{eqnarray}
where each energy subspace $\mathcal{H}_i\in\mathcal{H}$ has a detuning $\delta_i=E_i-E^\prime$ with respect to the qubit. We first consider the four-and five-qubit codes, and assume a scenario where $g_i=g=1$, and $\delta_1=0$, which subsequently fixes $\delta_i$ $\forall i$. While $H_{SA}$ can be constructed using any $\ket{\Psi_i}\in\mathcal{H}_i$ having the form as given in Eq.~(\ref{eq:archetype_states}), to keep the calculations simple, we choose an equal superposition, i.e.,  
$b_{i,\alpha}=1/\sqrt{d_i}$ $\forall \alpha$ for a specific $\mathcal{H}_i$. With BF noise applied to all qubits prepared in the logical state $\ket{\Psi_{0,0}}$ of the QECC, the success probability $P$ as functions of $\theta$ and $t$ are depicted in Figs.~\ref{fig:qubit_auxiliary}(a) and (c) for the four- and the five-qubit QECCs respectively. For the four-qubit code, with a fixed $\theta$, the variation of $P$ with $t$ exhibits regular revivals and collapses, while the regularity vanishes in the case of the five-qubit code, along with a reduction in the maximum success probability. Similar trend is observed for the case of the isotropic Heisenberg encoding, when $N$ is increased from $2$ to $4$ (see Figs.~\ref{fig:qubit_auxiliary}(e) and (g)) under the bit-flip noise. 

\begin{figure*}
    \centering
    \includegraphics[width=0.8\linewidth]{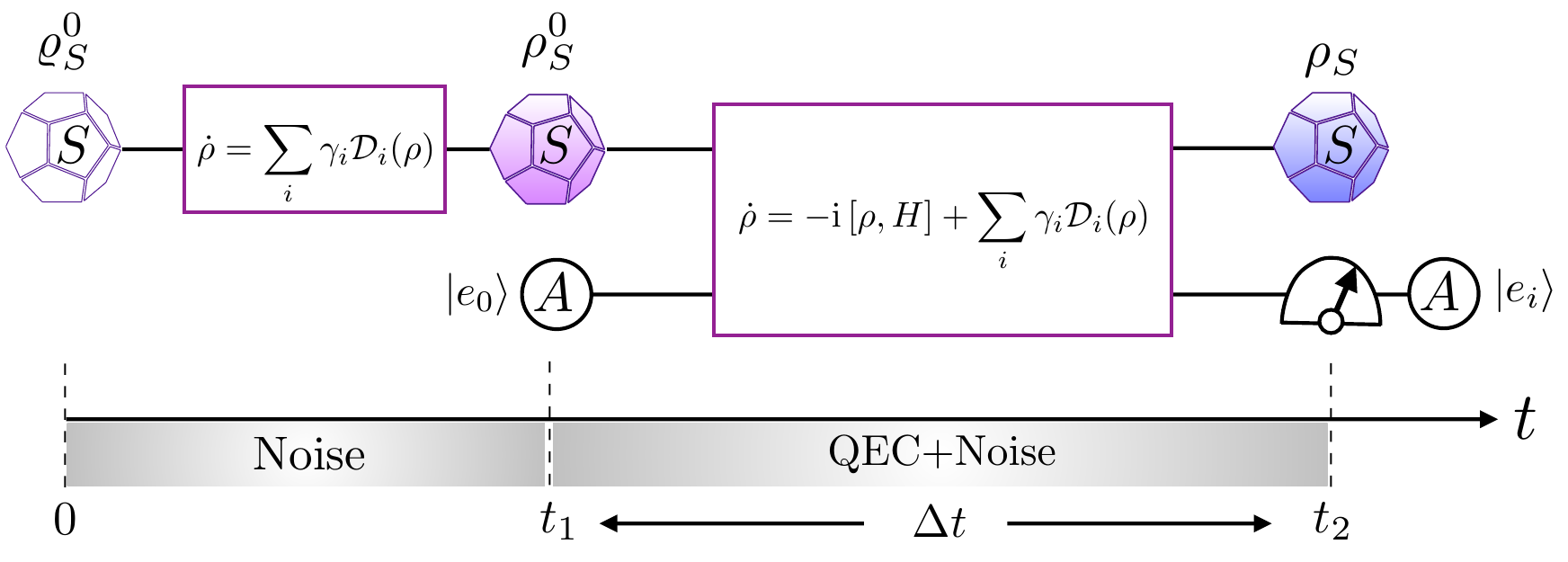}
    \caption{\textbf{QEC in the presence of resilient noise.} The effect of noise $\Lambda_0$ on the initial state $\varrho_S^0$ of the system $S$ is manifested in the state $\rho_S^0=\Lambda_0(\varrho_S^0)$. The QEC protocol starts at $t=0$ when the auxiliary system $A$, initialized in $\ket{e_0}$, is introduced, followed by the unitary evolution $U$ (Eq.~(\ref{eq:most_general_situation_closed})) that runs parallel to the noise $\Lambda$, so that the dynamics of $S$ is governed by the GKSL master equation (Eq.~(\ref{eq:continuous_QEC})). At $t=\Delta t$, a measurement on $A$ is performed, followed by a post-selection of the measurement outcome $e_i$ corresponding to the post-measured state $\ket{e_i}$ on $A$, so that the system collapses to $\rho_S$. If $\Lambda_0\equiv\Lambda$, the noise $\Lambda_0$ is assumed to act till $t=t_1$, followed by the QEC in the presence of noise $\Lambda$ till $t=t_2$, such that the duration of the QEC remains as $\Delta t=t_2-t_1$.}
    \label{fig:continuous_QEC}
\end{figure*}


\paragraph{Prescription 2.} In an alternative construction, we exploit the feature of the protocol discussed around Eq.~(\ref{eq:combined_subspaces}), and consider a single excited state subspace $\mathcal{H}_{1,2,\cdots,D}=\cup_{i=1}^D\mathcal{H}_i$, hosting an archetype state
\begin{eqnarray}
    \ket{\Psi_{1,2,\cdots,D}}=\sum_{i=1}^D\sum_{\alpha_i=0}^{d_i-1}b_{i,\alpha_i}\ket{E_{i,\alpha_i}}.
\end{eqnarray}
For simplicity, we consider 
\begin{eqnarray}
    \ket{\Psi_{1,2,\cdots,D}}=\frac{1}{\sqrt{D}}\sum_{i=1}^D\frac{1}{\sqrt{d_i}}\sum_{\alpha_i=0}^{d_i-1}\ket{E_{i,\alpha_i}},
\end{eqnarray}
for demonstration, and define (see Eq.~(\ref{eq:combined_subspace_interaction}))
\begin{eqnarray}
    H_{SA}&=&g(\ket{\Psi_{0,\gamma}e_1}\bra{\Psi_{1,2,\cdots,D} e_0}+\text{h.c.}).
\end{eqnarray}
Note that the reduction of $\mathcal{H}$ into two subspaces $\mathcal{H}_0$ and $\mathcal{H}_{12\cdots D}$ allows one to set $E_1=E^\prime$. The success rate obtained using this prescription is plotted as functions of $\theta$ and $t$ in Fig.~\ref{fig:qubit_auxiliary}. While the performance of Prescription 2 falls short in comparison to the same of Prescription 1 in the case of the four- and the five-qubit codes (see Figs.~\ref{fig:qubit_auxiliary}(b) and (d) for results of four- and five-qubit codes respectively using Prescription 2), in the case of the isotropic Heisenberg encoding (Figs.~\ref{fig:qubit_auxiliary}(f) and (h) for $N=2$ and $4$ respectively), Prescription 2 competes with Prescription 1. However, the feature of decreasing $P$ with increasing $N$ remains the same for both Prescriptions.

\section{QEC in presence of resilient noise}
\label{sec:resilient_noise}

The protocol discussed in Sec.~\ref{sec:Framework} assumes the application of the noise on the initial logical state, followed by a discontinuation of the noise when faulty state is subjected to the protocol. While this scenario finds its justification in situations where one has been able to successfully discard the source of noise and is now forced to work with the resulting noisy state, a more realistic assumption is to consider a situation where noise applies on the system
even during the application of the proposed protocol. This can be incorporated by considering an appropriate quantum master equation for the system instead of the unitary evolution alone, as discussed below (see Fig. \ref{fig:continuous_QEC}). 

Let us assume that the initial mixed state $\rho_S^0$ is the result of the noise represented by the Gorini-Kossakowski-Sudarshan-Lindblad (GKSL)  master equation~\cite{rivas2012,Lidar2019}
\begin{eqnarray}
    \frac{d\rho}{dt}=\sum_{i=1}^N\gamma_i\mathcal{D}_i(\rho),
\end{eqnarray}
where  $\mathcal{D}_i(.)$ is the superoperator describing the noise occurring with a rate $\gamma_i$ on the qubit $i$, and we have assumed independent local noise on each qubit. Starting from the initial state $\varrho_S^0$ at $t=0$, the noisy state $\rho_S^0$ is generated at $t=t_1$,  at which time  the QEC protocol as proposed in Sec.~\ref{sec:Framework} is applied. However, the  resilient local noise on each qubits in the system persists, such that the dynamics of the system is governed by the GKSL  master equation~\cite{rivas2012,Lidar2019} 
\begin{eqnarray}
    \frac{d\rho}{dt}=-\text{i}\left[H,\rho\right]+\sum_{i=1}^N\gamma_i\mathcal{D}_i(\rho),
    \label{eq:continuous_QEC}
\end{eqnarray}
with $H=H_S+H_A+H_{SA}$. The system-auxiliary duo is evolved according to Eq.~(\ref{eq:continuous_QEC}) till $t=t_2$, at which  the measurement on $A$ is performed and $\rho_S$ is obtained. We are interested in the duration of the QEC, which is given by $\Delta t = t_2-t_1$ (see Fig.~\ref{fig:continuous_QEC} for a schematic representation). The fidelity of $\varrho_S^0$ and $\rho_S$, denoted by $F$, can be computed using a suitable distance measure. In  our calculations, since $\varrho_S^0$ is pure ($\varrho_S^0=\ket{\Psi_{0,0}}\bra{\Psi_{0,0}}$, see Eq.~(\ref{eq:ground_states})), we compute \(F = \langle \Psi_{0,0} | \rho_S |\Psi_{0,0}\rangle \)~\cite{nielsen2010}. 

\begin{figure*}
   \centering
   \includegraphics[width=0.7\linewidth]{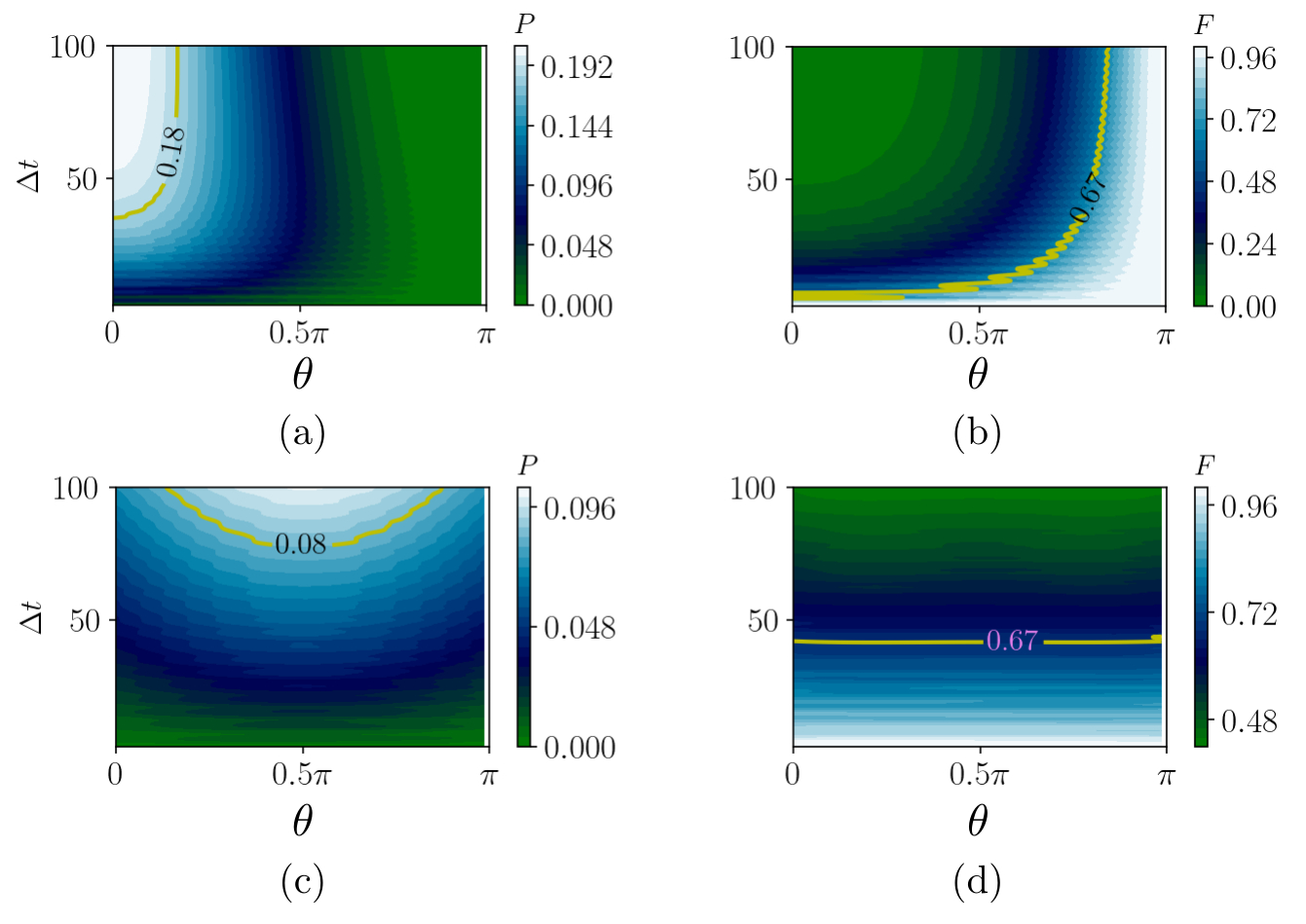}
   \caption{Performance of (a)-(b) three- and (c)-(d) four-qubit codes under the proposed protocol when amplitude damping noise acts for the entire period of QEC.  Map plot of (a), (c) probability \(P\) and (b), (d) fidelity $F$  with  \(\theta \) and \(\Delta t\).  The vertical axis is in the unit of \(g^{-1}\), while the horizontal axis is in radian.}
   \label{fig:resilient_noise}
\end{figure*}

In this paper, we consider  amplitude-damping noise acting on all qubits in the system, such that the GKSL master equation is given by~\cite{rivas2012,Lidar2019} 
\begin{eqnarray}
    \frac{d\rho}{dt}=-\text{i}\left[H,\rho\right]+\gamma\sum_{i=1}^N\bigg[\sigma^-_i\rho\sigma^+_i-\{\sigma^-_i\sigma^+_i,\rho\}\bigg],
\end{eqnarray}
where $\sigma^\pm=\sigma^x\pm\text{i}\sigma^y$, and we assume $\gamma$ to be identical for all qubits i.e., $\gamma_i=\gamma$ $\forall i=1,2,\cdots,N$.  The performances of the three- and four-qubit codes under persisting weak amplitude-damping noise  are demonstrated in Fig.~\ref{fig:resilient_noise}. It is clear from the figure that the effect of noise results in a decrease in $F$, while a low value of $P$ accompanies a high value of $F$ for the entire $(\theta,\Delta t)$ plane considered in this paper. We point out here that the design of the protocol requires a high population in the excited state manifold of the system (code) in order to provide a high value of $P$ (see Sec.~\ref{sec:Framework}). Since the noise is considered to be weak, a high value of $P$ is not observed.  This suggests that the adjustable parameter space may be tuned appropriately to achieve an optimal balance between the complementary behaviors of fidelity and success probability.


\section{Conclusions}
\label{sec:conclusion}

Quantum error correction (QEC) forms the backbone of fault-tolerant quantum computing by protecting fragile quantum information from decoherence. While numerous QEC schemes have been proposed, majority of the existing protocols are limited to correcting single-qubit errors distributed across multiple physical qubits.
In this paper, we developed a single-auxiliary assisted framework for QEC via purification that enables the correction of errors beyond a single qubit. In particular,  a quantum code initialized in a state resulting from noise in its logical subspace  is steered back to its logical subspace via a joint time-evolution of the system (code)-auxiliary duo, followed by a measurement and a post-selection of the measurement outcome on the auxiliary. We demonstrated the success of the protocol in  the case of low-distance non-CSS codes such as the three-, four-, and five-qubit codes as well as isotropic Heisenberg encoding of a small number of qubits. Remarkably, we showed that the 
the logical subspace can be successfully restored even under noise types that the codes are not conventionally designed to correct, thereby highlighting the power of the protocol. We further explored the performance of the protocol in the practical scenario  involving resilient noise, where the noise continued to  affect  the system throughout the correction process.

Our work provides an alternative approach for achieving QEC for quantum systems via a purification procedure using only a single auxiliary. In this respect, it connects naturally to recent explorations of auxiliary-assisted cooling protocols~\cite{Yan2021,Yan2022,Konar2022,Ghosh2024_arX}, opening avenues for interpreting the proposed method from thermodynamic and device-engineering perspectives. It is also in alignment with the explorations of measurement-based QEC schemes~\cite{hillmann_2025},  and can also be looked at as a ground-state preparation protocol using cooling~\cite{Motlagh2024_arX}.

Our work assumes a number of ideal scenarios,  including perfect initialization of the auxiliary in its ground state, ideal projection measurements, and flawless implementation of the system-auxiliary unitary. However, in reality, 
imperfections in these components are inevitable, making it essential to examine the robustness of the protocol under such deviations. Further, while the protocol can, in principle, be demonstrated for quantum systems (codes) of arbitrary sizes, accessing the full spectrum of the Hamiltonian in order to design the interaction Hamiltonian involving system-auxiliary pair is always challenging.   
This motivates the development of alternative \emph{scalable} strategies to design the interaction Hamiltonian utilizing appropriate information of the Hilbert space.

\acknowledgements 

The authors acknowledge the cluster computing facility at Harish-Chandra Research Institute and the use of \href{https://github.com/titaschanda/QIClib}{QIClib} -- a modern C++ library for general purpose quantum information processing and quantum computing. This research was supported in part by the ``INFOSYS scholarship for senior students''. A.K.P and A.S.D acknowledge the support from the Anusandhan National Research Foundation (ANRF) of the Department of Science and Technology (DST), India, through the Core Research Grant (CRG) (File No. CRG/2023/001217, Sanction Date 16 May 2024). A.S.D. acknowledges support from the project entitled "Technology Vertical - Quantum Communication'' under the National Quantum Mission of the Department of Science and Technology (DST)  ( Sanction Order No. DST/QTC/NQM/QComm/2024/2 (G)).

\bibliography{ref}

\newpage   

\onecolumngrid

\appendix

\end{document}